\documentclass{article}

\usepackage{titlesec}
\usepackage{arxiv}

\usepackage[utf8]{inputenc} % allow utf-8 input
\usepackage[T1]{fontenc}    % use 8-bit T1 fonts
\usepackage{hyperref}       % hyperlinks
\usepackage{url}            % simple URL typesetting
\usepackage{booktabs}       % professional-quality tables
\usepackage{amsfonts}       % blackboard math symbols
\usepackage{nicefrac}       % compact symbols for 1/2, etc.
\usepackage{microtype}      % microtypography
\usepackage{lipsum}		% Can be removed after putting your text content
\usepackage{graphicx}
\usepackage{natbib}
\usepackage{doi}

\usepackage[nolist]{acronym}
\usepackage{tabularx}
\usepackage{subcaption}
\usepackage{caption}
\usepackage{capt-of}
\usepackage{subcaption}
\usepackage{siunitx}
\usepackage{amsmath}
\usepackage{cleveref}
\usepackage{comment}

\begin{acronym}
    \acro{KIT}{Karlsruhe Institute of Technology}
    \acro{FZJ}{Forschungszentrum Jülich}
    \acro{DH}{District Heating}
    \acro{TABULA}{Typology Approach for Building Stock Energy Assessment}
    \acro{DHW}{Domestic Hot Water}
    \acro{HIU}{Heat Interface Unit}
    \acro{HVAC}{Heating, Ventilation and Air Conditioning}
    \acro{FMI}{Functional Mock-up Interface}
    \acro{FMU}{Functional Mock-up Unit}
    \acro{LPG}{LoadProfileGenerator}
    \acro{BEMs}{Building Energy Models}
    \acro{HiFi}{High-Fidelity}
    \acro{LoFi}{Low-Fidelity}
    \acro{IBPSA}{International Building Performance Simulation Association}
    \acro{BIM}{Building Information Modeling}
    \acro{ASHRAE}{American Society of Heating, Refrigerating and Air-Conditioning Engineers}
    \acro{UFH}{Underfloor Heating}
    \acro{median RMSE}{Median Root Mean Square Error}
    \acro{OSM}{OpenStreetMap}
    \acro{GUI}{Graphical User Interface}
    \acro{ROM}{Reduced Order Model}
    \acro{MPC}{Model Predictive Control}
    \acro{SVG}{Scalable Vector Graphics}
\end{acronym}

\title{JanusBM: A Dual-Fidelity Multi-Zone White-Box Building Modeling Framework}

\author{ \href{https://orcid.org/0009-0005-0067-8019}{\includegraphics[scale=0.06]{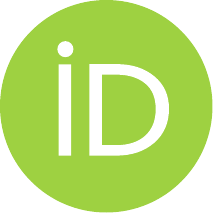}\hspace{1mm}Haozhen~Cheng}\\
	Karlsruhe Institute of Technology\\
	76344 Eggenstein-Leopoldshafen\\
	\texttt{haozhen.cheng@kit.edu}\\
	\And
	\href{https://orcid.org/0000-0002-1463-7606}{\includegraphics[scale=0.06]{orcid.pdf}\hspace{1mm}Hüseyin K.~Çakmak} \\
	Karlsruhe Institute of Technology\\
	76344 Eggenstein-Leopoldshafen\\
	\texttt{hueseyin.cakmak@kit.edu}\\
    \And
	\href{https://orcid.org/0000-0002-3572-9083}{\includegraphics[scale=0.06]{orcid.pdf}\hspace{1mm}Veit~Hagenmeyer} \\
	Karlsruhe Institute of Technology\\
	76344 Eggenstein-Leopoldshafen\\
	\texttt{veit.hagenmeyer@kit.edu}\\
}

\renewcommand{\shorttitle}{A Dual-Fidelity Multi-Zone White-Box Building Modeling Framework}

\hypersetup{
pdftitle={JanusBM: A Dual-Fidelity Multi-Zone White-Box Building Modeling Framework},
pdfsubject={eess.SY},
pdfauthor={Haozhen~Cheng, Hüseyin K.~Çakmak, Veit~Hagenmeyer},
pdfkeywords={Building, Multi-zone, White-box, Modelica, Modeling, Simulation, Hydronic, Validation, Calibration},
}
\begin{document}
\maketitle

\begin{figure}[h]
  \centering
  \includegraphics[width=0.95\textwidth]{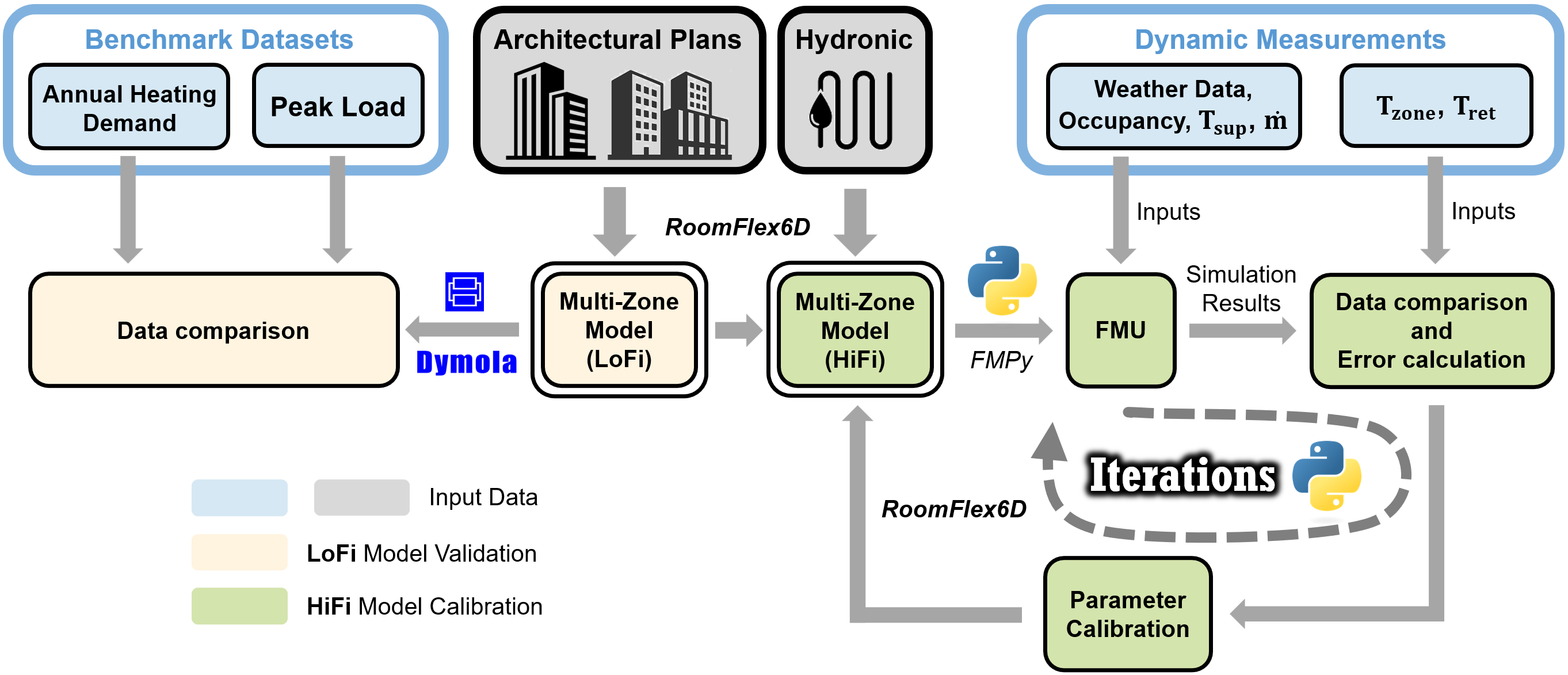}
  \caption{Overview of the structured dual-fidelity modeling framework JanusBM, including data input, model generation, validation, and iterative calibration workflow.}
  \label{fig: methodology}
\end{figure}

\vspace{0.7em}

\begin{abstract}

	Accurate building energy models are crucial for analyzing sector-coupled energy systems, where buildings interact with electrified heating, energy storage, and advanced control across various scenarios. High-fidelity (HiFi) white-box models that resolve hydronic distribution and emitter dynamics can capture short-term transients, yet their numerical stiffness and computational burden limit long-term simulations and large-scale scenario exploration. Conversely, reduced-order low-fidelity (LoFi) representations enable rapid annual assessments but may fail to capture the hydronic- and control-induced dynamics that govern transient and peak behavior. This paper proposes a dual-fidelity, multi-zone white-box building modeling framework, which is called JanusBM, built on a novel topology-driven modeling tool RoomFlex6D, coupling a HiFi hydronic model and a LoFi ideal-load surrogate that removes explicit hydronic states in Modelica. To ensure applicability and physical consistency across time scales, we introduce a two-stage hybrid validation and calibration pipeline that uses complementary data: the IEA EBC Annex 60 benchmark for energy-scale validation and time-series measurements from real-world experimental buildings for hydronic dynamics-scale calibration. Results show that the generated LoFi models achieve a high degree of consistency with Annex 60 benchmark on the energy scale, and the proposed calibration workflow robustly improves loop-level return water temperature transients and zone-level temperature dynamics. Moreover, the LoFi model achieves orders-of-magnitude faster simulations suited to annual energy analyses, whereas the HiFi model becomes necessary when the required heat differs from the actual delivered heat due to distribution and control limitations, especially in transient and peak-oriented assessments.
\end{abstract}

% keywords can be removed
\keywords{Building \and Multi-zone \and White-box \and Modelica \and Modeling \and Simulation \and Hydronic \and Validation \and Calibration}

\section{Introduction}

    As a central element in sector-coupled energy systems, buildings serve as the primary interface between various energy infrastructures, including heat and electricity \citep{Gea2021}. As electrified heating technologies (e.g., heat pumps), distributed photovoltaics, energy storage, and flexible loads become increasingly prevalent, system-level studies are no longer limited to estimating annual energy demand \citep{Jing2022, Rinaldi2022}; instead, they must evaluate buildings across a wide range of design and operational scenarios, where short-term dynamics and operational constraints can play a decisive role \citep{Drgona2020, Rinaldi2021, Vera2024}.
    
    \ac{HiFi} multi-zone white-box models, typically implemented in physics-based environments such as EnergyPlus \citep{Crawley2001} and TRNSYS \citep{Klein1988}, that explicitly characterize building envelopes, thermal mass, and hydronic heating systems, can reproduce transient behavior including water-side dynamics, emitter inertia, and their interaction with zone temperatures. However, as zoning resolution increases, the hydronic networks and control loops introduce numerical stiffness and rapidly increase the system dimension, making simulation costly and sensitive to solver settings \citep{Blum2021}. Consequently, \ac{HiFi} representations are often impractical for long-term simulations, large-scale scenario scans, or optimization workflows that require thousands of evaluations \citep{Bhattacharya2020}. 
    
    Conversely, reduced-order or \ac{LoFi} models enable fast annual simulations and are widely used for convenient parameter configuration of \ac{BEMs} and long-term planning studies, but they often neglect hydrodynamics constraints and detailed control interactions. This simplification can systematically fail to represent short-term transients and peak-oriented indices that are critical in sector-coupled applications, thereby subsequent conclusions regarding scale margin, flexibility assessments, and controller robustness can be biased \citep{ISO2017}. To accurately represent building systems, multi-zone \ac{BEMs} have been extensively studied and are becoming increasingly mature. However, researchers often either over-refine zoning in expensive \ac{HiFi} environments, making large-scale scenario exploration infeasible, or oversimplify accuracy for speed, leading to systematic biases in peak-oriented conclusions and incorrect controller evaluations.

\subsection{Contributions}
    To make model-fidelity choices explicit and comparable, we propose the unified dual-fidelity, multi-zone white-box building modeling framework JanusBM, which is shown in Figure~\ref{fig: methodology}, enabling controlled and quantitative comparisons between zoning resolution, model fidelity, and computational cost under consistent envelope assumptions. Specifically, we

    \begin{itemize}
        \item introduce the topology-driven modeling tool RoomFlex6D, to automatically generate zoning-consistent modular multi-zone envelope models, and to configure hydronic loop topologies and their couplings to zones in a systematic behavior;
        \item integrate two complementary representations \ac{HiFi} and \ac{LoFi} derived from the same zoning and envelope description into the proposed dual-fidelity framework JanusBM, ensuring that fidelity comparisons are not affected by envelope modeling differences;
        \item develope a two-step validation-calibration workflow for the proposed framework using complementary datasets, to validate the \ac{LoFi} model at energy scale, and to calibrate the \ac{HiFi} model in terms of dynamic performance using a closed-loop RoomFlex6D–\ac{FMU}–Python workflow; and
        \item demonstrate, how fidelity choices impact computational burden, thereby providing practical evidence to guide fidelity selection in sector-coupled studies.
    \end{itemize}
    
\subsection{Related Work}
    
    Multi-zone white-box building energy modeling is a mature research area. Compared with single-zone representations, multi-zone models indicate inter-zonal heat transfer paths and system-building interactions, such as solar and internal gains, and zone-specific \ac{HVAC} interactions, which are essential for capturing dynamic responses, peak loads, and system dynamics under hydronic heating operation \citep{Zhang2022}. Existing studies related to this topic can be broadly grouped into four categories: (a) automatic zoning strategies, that balance zoning resolution against computational burden \citep{Jansen2022, Shin2022}, (b) \ac{LoFi} or reduced-order modeling toolchains, including ideal-load surrogates and \ac{ROM}-based workflows, which enable fast simulations for large-scale simulations for urban planning and scenario analysis \citep{Belic2021, Jansen2022}, (c) interoperability and standardized Modelica-based workflows, where detailed libraries establish modeling conventions and support \ac{BIM}-to-Modelica transformations \citep{Wetter2021, Wetter2019} and \ac{FMI}-based co-simulation \citep{WetterSpawn2021} for multiphysics building and \ac{HVAC} modeling, and (d) validation and calibration practices, in which benchmark datasets and real-world measurements \citep{Langner2025} provide complementary---yet asymmetric---evidence of model credibility across different scales \citep{Ohlsson2021}.

    \textbf{Zoning strategy and structured simplification.}
    Numerous studies have investigated zoning definition and zoning strategies based on geometry, usage patterns, and facade characteristics, as well as structured simplification methods. \citet{Georgescu2015} proposed a structured model simplification technique that couples zoning structure with simulation accuracy and complexity, arguing that physically structured zoning is preferable to arbitrary subdivision. \citet{Chen2018} introduced a three-zone approach guided by ASHRAE~90.1 Appendix G guidelines \citep{ashrae2026}, combined with floor or zone multipliers, to accelerate simulation while maintaining acceptable accuracy for annual assessments. \citet{Shin2019} provided a comprehensive review of zoning methods and highlighted the recurring trade-off between zoning resolution and computational burden. In practical building energy system studies, however, building geometry often varies from project to project, and the simulation workflow requires repeated model generation under different zoning schemes, envelope configurations, and system layouts. Traditional geometry-driven methods based on detailed 3D model libraries or \ac{BIM} data may not be flexible enough in this context, as even minor changes in zoning or topology often require significant manual rework and consistency checks. Another related modeling framework, ALICE/ALICE2Modelica, was developed by \citet{Mork2024} to support automated generation of large-scale, multi-zone building models. In its current workflow, building orientation, adjacency relationships, and interior/exterior wall identification are derived from per-floor \ac{SVG} plans. This dependency means that changes in zoning or building layout typically may require manual updates to the floor-plan inputs, limiting rapid re-zoning and structural reconfiguration. Motivated by this limitation, a topology-driven Lego\textregistered-like modeling tool RoomFlex6D is introduced by the present paper, enabling more flexible and reconfigurable multi-zone model construction.

    \textbf{Reduced-order toolchains.}
    To support large-scale scenario exploration and planning studies, many approaches employ reduced-order or \ac{LoFi} representations, including ideal-load surrogates and \ac{ROM}-based workflows used in urban simulations. These methods are highly scalable, and are convenient for parameter configuration and long-term analyses. However, early work in this direction---for example, TEASER \citep{Remmen2018} used by Modelica AixLib, and CityBES \citep{Chen2017}---already recognized that single-zone or lumped models may fail to accurately reflect the spatial distribution of temperature and loads, particularly when facade exposure, occupancy patterns, and equipment operation differ across zones, thereby limiting the credibility of both comfort assessment and control-oriented analysis \citep{Johari2020}.

    \textbf{Modelica interoperability workflows.}
    The Modelica-based building and \ac{HVAC} libraries, including \ac{IBPSA}-related \citep{ibpsa2026} Modelica libraries \citep{Jorissen2018, Maier2023, Nytsch2012, Wetter2014}, provide a robust foundation for multiphysics, multiscale simulations, enabling explicit representation of building envelopes, thermal mass, fluid networks, and control systems. In parallel, interoperability efforts such as the \ac{IBPSA} activity Annex 60 \citep{Wetter2017} demonstrated the potential of modular, tool-independent simulation chains. Nevertheless, these efforts rarely support controlled cross-fidelity comparisons, making it difficult to validate the model correctness across time scales and accuracy levels.

    \textbf{Validation and calibration across time/accuracy scales.}
    A further obstacle is the asymmetry in the availability of data for validation and calibration: benchmark datasets typically offer standardized annual indicators but lack time-resolved measurements, while measured time-series data \citep{Italo2026, erfan2025} provide rich, detailed signals but are rarely suitable for reliable annual energy calibration under controlled conditions \citep{Ohlsson2021}. This mismatch becomes particularly critical in the context of \textbf{digital twins}, which are not just standalone models, but also executable representations of as-built assets that are continuously informed by an evolving stream of operational data, and require mechanisms to dynamically update model parameters to remain consistent with their physical counterparts over time \citep{Wright2020}. Consequently, establishing a credible building digital twin necessitates systematic model calibration rather than one-off validation: energy-scale benchmarks can constrain long-term consistency, while dynamics-scale measurements enable parameter updating for short-term behaviors and peak conditions.

\subsection{Outline}
    Building on the identified research gaps, this paper proposes a dual-fidelity white-box building modeling framework, including the topology-driven RoomFlex6D tool and its Modelica-based generation of \ac{LoFi} and \ac{HiFi} model instances under consistent envelope assumptions, as presented in Section~\ref{sec: model construction}. Section~\ref{sec: V&C pipeline} details the two-stage validation–calibration pipeline, while Section~\ref{sec: zoning} introduces the zoning strategy and explains how zoning configurations are compared in a controlled manner. Section~\ref{sec: Evaluation} reports the evaluation results, covering \ac{LoFi} validation and \ac{HiFi} calibration, together with the resulting accuracy–runtime trade-off, and finally, Section~\ref{sec: Conclusion and outlook} concludes the paper and outlines future work.

\section{Model Construction}
\label{sec: model construction}
\begin{comment}
    A structured dual-fidelity modeling framework for multi-zone buildings will be introduced in this section. First, the novel workflow RoomFlex6D is described, including its parametric representation of zones and automated multi-zone model generation capabilities. Then, dual-fidelity approaches are introduced. Lastly, a validation and calibration process utilizing complementary benchmark and real-world experimental datasets is introduced along with the corresponding zoning strategy.
\end{comment} 
    
    A novel Lego\textregistered-like multi-zone white-box building modeling tool, that can generate both \ac{HiFi} and \ac{LoFi} models with the same building envelope configuration, is introduced in the Sections below.
    
\subsection{RoomFlex6D Modeling Tool}

    To describe building instances in a way that directly supports rapid multi-zone model generation and systematic zoning studies, the present paper proposes a novel, lightweight, and simulation-oriented modeling tool, RoomFlex6D. Unlike general semantic building standards such as CityGML \citep{CityGML2026}, IFC \citep{IFC2026}, or gbXML\citep{gbXML2026}, which enable rich geometric and semantic interoperability across domains, RoomFlex6D can generate a minimally sufficient description for thermal simulation that can be edited and repeatedly regenerated, and is highly comparable across different zoning schemes. Specifically, RoomFlex6D encodes the building as a topology-driven, face-level adjacency graph (zone, orientation, boundary type, and neighbor) that automatically maintains parameter consistency between adjacent faces of adjacent thermal zones. 
    
    Furthermore, the presented tool allows for the flexible construction of hydronic loop topologies and their coupling with zones and embedded/non-embedded heating elements, which are crucial for transient hydronic dynamics and controller interactions, which cannot be implemented in a stable and standardized manner in typical semantic models \citep{Luo2021}.
\begin{comment}
    Importantly, this workflow supports the generation of multi-zone models with two levels of fidelity from the same underlying zoning and envelope description: a \ac{HiFi} model with an explicit representation of the hydronic system, and a reduced-order \ac{LoFi} model that removes hydronic dynamics but retains the same zoning and thermal structure.
\end{comment}

\begin{figure}[t]
      \centering
      \includegraphics[width=0.6\linewidth]{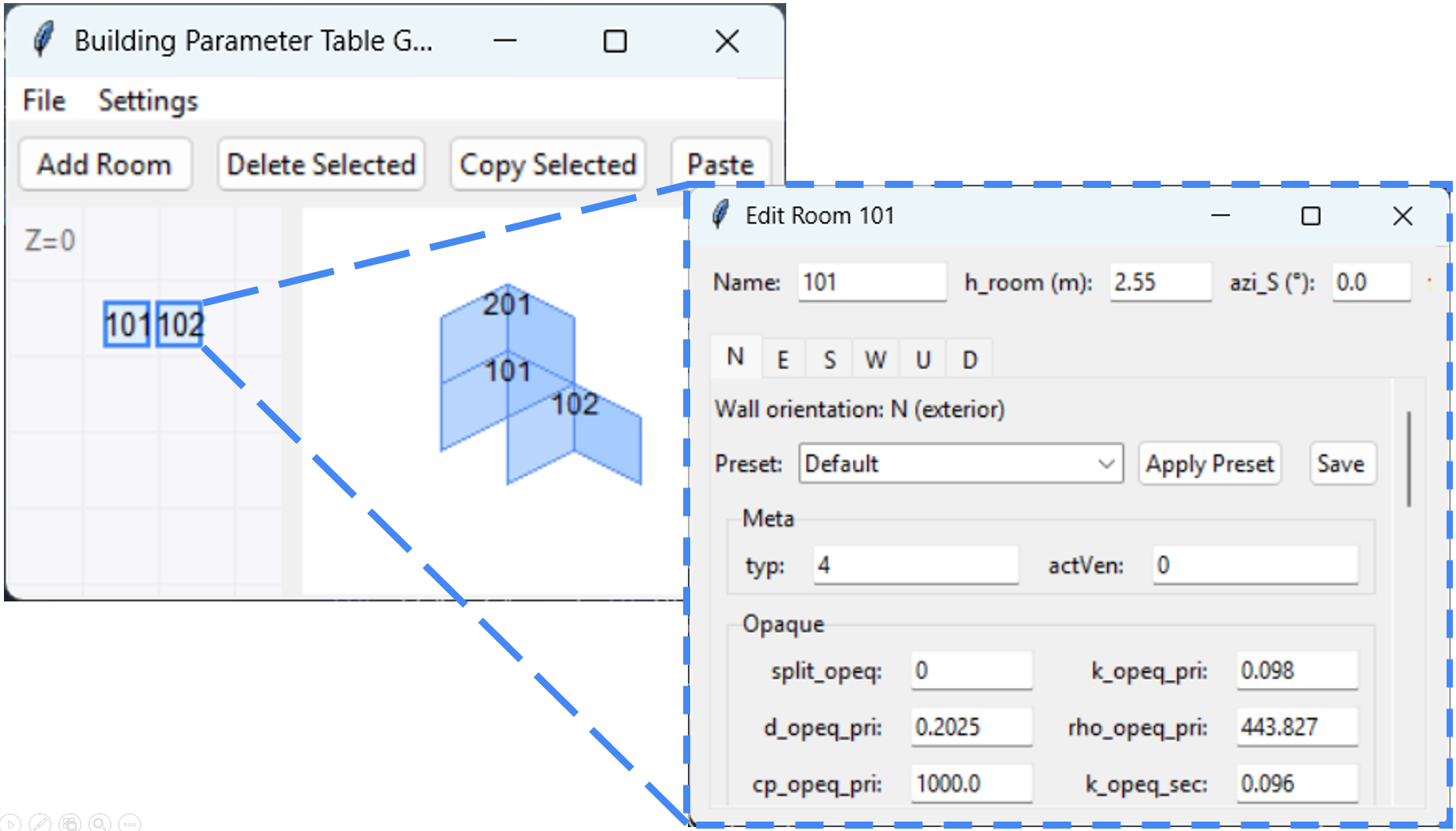}
      \caption{RoomFlex6D GUI for topology-driven zoning and face-level boundary assignment: drag-and-drop zone layout (left), per-zone face properties and hydronic mapping (right), exporting a reproducible parameter table. See Appendix~\ref{sec: appendix_GUI} for more details.}
      \label{fig: GUI}  
    \end{figure}

\begin{figure}[h]
          \centering
          \includegraphics[width=0.65\linewidth]{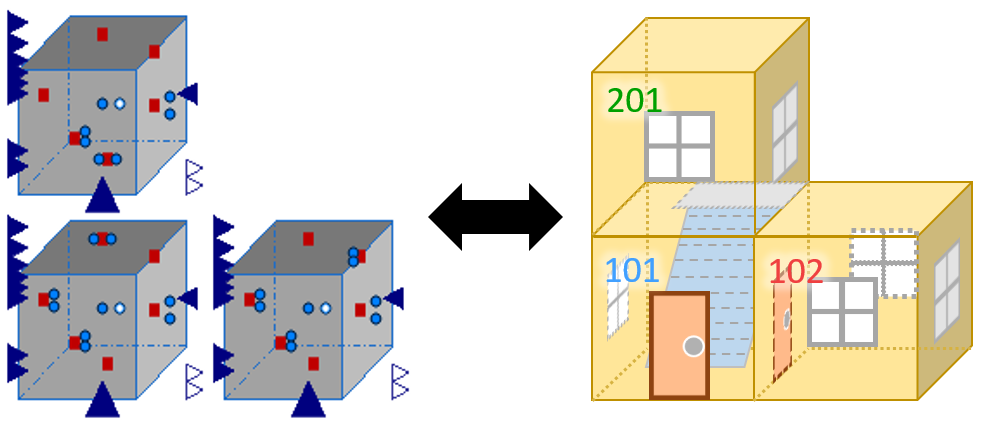}
          \caption{Topology-to-Modelica translation example: a 3-zone model instance generated from the RoomFlex6D parameter table (left) and its schematic graph representation (right).}
          \label{fig: example mo building}  
    \end{figure}

    RoomFlex6D provides a Python-based topology-driven \ac{GUI} in Figure~\ref{fig: GUI} that encodes a multi-zone building as a set of orthogonal zone volumes, and exports a face-level parameter table for automated Modelica model generation. Each zone is connected to its neighbors (or the ambient) through six directional faces (N/E/S/W/U/D), where each face is assigned an adjacent-face boundary type (Table~\ref{tab: adj types} based on \citet{Wetter2006}). This structured Lego\textregistered-like representation ensures the uniqueness of adjacency relationships while avoiding ambiguous or non-manifold connections, enabling controlled zoning changes while keeping envelope assumptions comparable across model instances. The exported parameter table serves as the source for generating and updating Modelica models (see Figure~\ref{fig: example mo building} for an example). Based on the generated structural parameter table (Table~\ref{tab: roomflex params}), the heat transfer mechanisms achievable through the RoomFlex6D modeling tool include:
    \begin{itemize}
    \item conductive heat transfer from walls, floors, and roofs,
    \item convective heat exchange between zoned air and interior surfaces,
    \item convective heat exchange on exterior surfaces affected by external temperature and wind speed,
    \item solar radiation through transparent components, and the heating of the external surface by solar radiation,
    \item internal heat generated by people and equipment, and
    \item heat loss due to infiltration and ventilation.
    \end{itemize}

    \begin{table}[t]
    \centering
    \caption{Face boundary-type codes used in RoomFlex6D to parameterize openings/ground coupling for each zone face.}
    \label{tab: adj types}
    \begin{tabularx}{0.7\linewidth}{cX}
    \hline
    Boundary Type & Comments \\
    \hline
    1 & There is an opening that cannot be closed. \\
    2 & There is an operable window(s). \\
    3 & There is an operable door(s). \\
    4 & There is no operable opening, and the exterior is not ground. \\
    5 & There is no operable opening, and the exterior is ground. \\
    \hline
    \end{tabularx}
    \end{table}

    \begin{table}[t]
        \centering
        \caption{Face-level parameter-table minimal schema enabling zoning-consistent envelope and hydronic coupling (each row = one zone face). Full list in Appendix~\ref{sec: appendix_Parameter table}.}
        \label{tab: roomflex params}
        \renewcommand{\arraystretch}{1.1}
        \begin{tabularx}{0.5\linewidth}{l c}
        \hline
        \textbf{Parameter} 
        & \textbf{Symbol} \\
        \hline
        Primary zone ID      
        & $z_{\mathrm{pri}}$ \\
        
        Face orientation     
        & $ori$ \\
        
        Adjacent zone ID     
        & $z_{\mathrm{adj}}$ \\
        
        Boundary type        
        & $typ$ \\
        
        Ventilation flag     
        & $actVen$ \\
        
        Split flag   
        & $split$ \\
        
        Thermal parameter  
        & ${k,d,\rho,c_p,A}_{\mathrm{opeq,fra,gla,open,RS}}$ \\
    
        Hydronic  
        & $heat_{\mathrm{type,loop,order,ctrl}}$ \\
    
        Building orientation
        & $azi_{\mathrm{S}}$ \\
    
        \hline
        \end{tabularx}
    \end{table}

\subsection{\ac{HiFi} Hydronic Heating System Model}

    As described in Table~\ref{tab: roomflex params}, in addition to the envelope and multi-zone structure, RoomFlex6D supports an explicit representation of the building heating hydronic and its coupling to zone heat transfer. Typical configurations such as radiator heating and \ac{UFH} are modeled as dynamic heaters with finite thermal capacity, exchanging heat with zone air (or embedded constructions) via convection and radiation. The hydronic network comprises supply/return piping, hydraulic resistances, circulation pumps, and control elements (e.g., control and mixing valves), enabling simulation of operational dynamics, such as switching cycles and transient responses to boundary-condition changes.

\begin{figure}[t]
      \centering
      \includegraphics[width=0.55\linewidth]{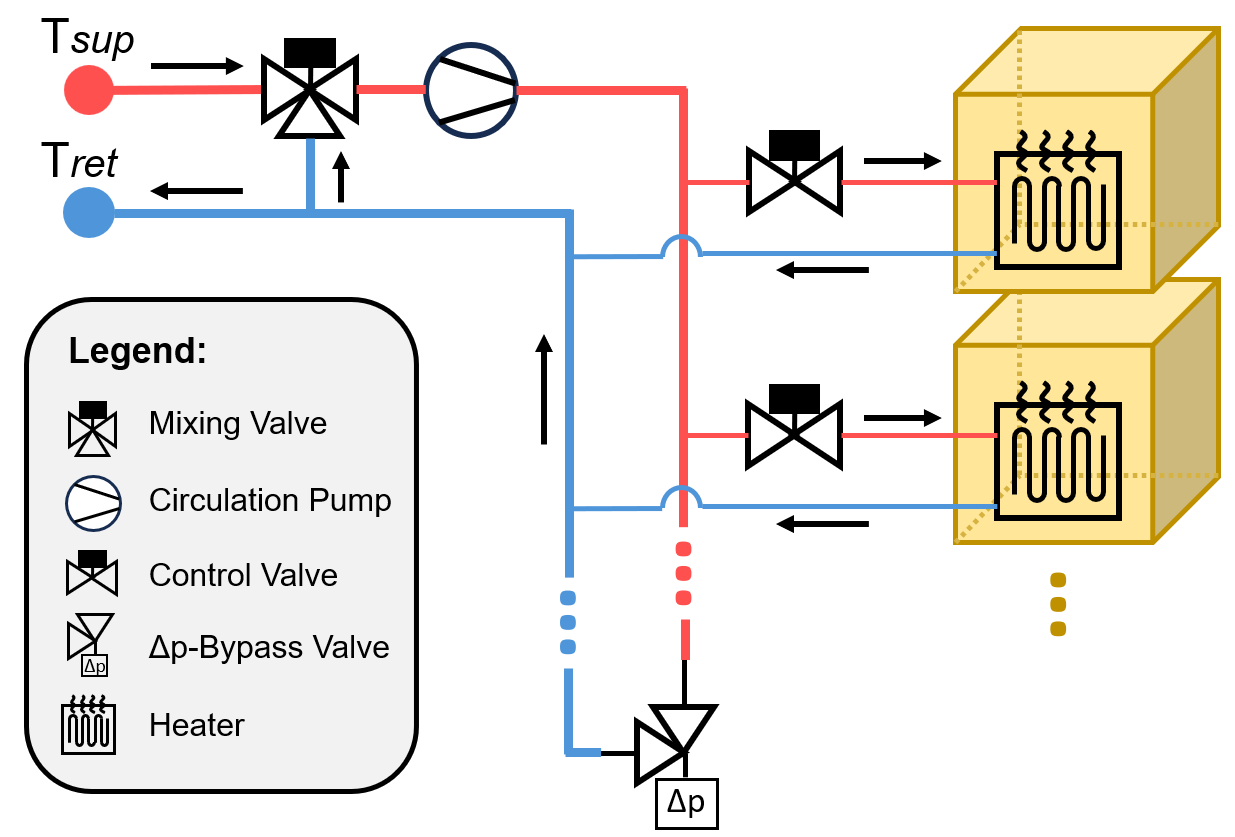}
      \caption{HiFi model structure generated by RoomFlex6D: explicit hydronic network, emitters, and zone coupling: source of stiffness but required for transient loop dynamics.}
      \label{fig: hydronic}  
    \end{figure}

    Based on the generated zone structure, the \ac{HiFi} model integrates these hydronic components with multi-zone heat transfer, solar gains, internal loads, infiltration, and building thermal mass, as shown in Figure~\ref{fig: hydronic}. However, this coupling introduces numerical stiffness due to multiple timescales --- from rapid hydronic control to the slow thermal response of building structures. As the number of thermal zones increases, the system dimensionality grows rapidly, leading to increased computational costs and greater sensitivity to solver settings (e.g., time step and tolerance). This motivates the complementary \ac{LoFi} ideal-load model used for large-scale and year-round studies within the JanusBM framework.

\subsection{\ac{LoFi} Ideal-Load Building Model}

    To enable rapid simulations suitable for year-round analysis and large-scale scenario exploration, a reduced-order ideal-load multi-zone building \ac{LoFi} model is derived as a surrogate representation of the \ac{HiFi} model. This model retains the same multi-zone structure, envelope description, and boundary conditions generated by RoomFlex6D, while removing explicit representations of hydronic components and fluid dynamics. It provides heat to each zone via an idealized heat input that meets the heating requirements to maintain a preset zone temperature. Therefore, the \ac{LoFi} model analyzes the thermodynamics of the building envelope and zone air volumes while simplifying the behavior of the internal hydronic system. Consequently, the solver burden caused by rapid hydronic timescales is eliminated, which is a major source of numerical stiffness in \ac{HiFi} models.
    
    Despite the simplifications, key physical properties relevant to system-level analysis are preserved in the \ac{LoFi} model, including explicit simulations of multi-zone heat transfer, solar radiation gain, internal loads, infiltration, and building thermal mass, thus capturing the overall building thermal behavior with reasonable accuracy. This makes the simplified model particularly suitable for year-round simulations, extensive parameter scans, and scenario sets involving multiple energy system configurations, to support tasks such as annual energy assessments, scenario comparisons, and preliminary system design.
    
    In summary, within the proposed dual-fidelity framework, the simplified ideal load model is not intended to replace the full hydronic model. Instead, it serves as a computationally efficient alternative model that can be calibrated and validated against high-fidelity models and benchmark data.

\section{Validation and Calibration Pipeline}
\label{sec: V&C pipeline}
    To ensure physical consistency and applicability across different model accuracies, the present paper introduces a structured validation and calibration process. 
    
    \textbf{The Annex 60 benchmark dataset} provides standardized building descriptions and annual heating demand references, enabling energy-scale validation under well-defined boundary conditions \citep{Wetter2017}. However, Annex 60 does not provide time-resolved measurements, thus failing to support validation of transient system behavior.
    
    \textbf{The experimental building dataset} from the Living Lab at \ac{KIT} Campus North is available \citep{erfan2025}. Conversely, this dataset contains local weather data, time-series measurements of zone temperature, supply and return water temperatures, and mass flow rates in each of the hydronic heating loops (\ac{UFH}), and is stored in a time series database with a Grafana (grafana.com) interface for further data analysis and export \citep{hagenmeyer2016}. While these measurements capture detailed dynamic behavior, the experimental building operates under user-driven conditions and cannot provide reliable annual heating demand values suitable for direct energy-scale calibration.
    
    This asymmetry motivates a two-step validation and calibration strategy described below, leveraging the strengths of each dataset while avoiding their respective limitations.

    \textbf{Workflow of Pipeline.}
    In the present paper, these two types of data are used by the \ac{LoFi} and \ac{HiFi} models, respectively, to validate the model reliability at energy scales, and to calibrate model parameters at dynamic scales by using a novel iterative calibration workflow, as shown in the bottom right corner of Figure~\ref{fig: methodology}. It outlines the closed-loop parameter calibration process based on an iterative RoomFlex6D–\ac{FMU}–Python workflow:
    \begin{itemize}
        \item \textbf{data input:} Building plans and geographic information (such as architectural drawings or \ac{OSM} \citep{OSM2026}) serve as structured inputs, and during simulation, the model receives external boundary and system inputs from Living Lab experimental data \citep{erfan2025}, including weather data, occupancy-related signals, and \ac{UFH} information;
        \item \textbf{model generation and simulation:} For the \ac{HiFi} models calibration, RoomFlex6D automatically generates multi-zone white-box building models from the inputs mentioned above, which are then exported as executable \ac{FMU}s, and the Python package FMPy handles batch simulation iterations;
        \item \textbf{loss calculation and parameter calibration:} The simulation outputs are aligned and compared with synchronously measured data from the Living Lab as time series to calculate errors, and construct a calibration objective function, which drives the parameter update module to iteratively correct the parameters to be calibrated, and then the updated parameters are written back to the model generation-simulation link for the next round of simulation. 
    \end{itemize}

    \textbf{Algorithm: }Specifically, within a selected calibration time window, the observed signal set is defined as \(\mathcal{S}=\{T_{\mathrm{zone}},\,T_{\mathrm{ret}}\}\), where \(T_{\mathrm{zone}}\) the zone air temperature and \(T_{\mathrm{ret}}\) denotes the return water temperature of each hydronic heating loop. The parameter vector to be calibrated is \(\Phi=\{dis_{\mathrm{pip}},\,h_{\mathrm{int}}\}\), where $dis_{\mathrm{pip}}$ represents the pipe spacing distance of the embedded floor-heating emitters and $h_{\mathrm{int}}$ denotes the effective internal convective heat-transfer coefficient between the floor-heating surface and the surrounding zone air. In the implementation, $dis_{\mathrm{pip}}$ is treated as a global (shared) parameter across all loops to meet real-world physics, while $h_{\mathrm{int}}$ is updated per loop. Both are optimized in log-space to enforce positivity and improve numerical stability.

    \begin{itemize}
        \item \textbf{Flow-gating:} For each iteration, simulation outputs are aligned to the measurement time grid within the calibration time window. To prevent non-informative segments from distorting the calibration results, the return water channels are evaluated only during time steps with sufficient measured mass flow rate, i.e., \(\dot{m}(t)\ge \dot{m}_{\min}\), which means that periods with \(\dot{m}\approx 0\) do not contribute to the return temperature residuals. This is because in the one-dimensional fluid network model of Modelica, the convection processes become numerically ill-conditioned when $\dot{m}$ approaches 0, and temperature states within volume elements can be affected by fluid mixing, leading to distorted return temperature signals that fail to represent the true heat transfer physics.
        \item \textbf{Bias-driven update:} The scalar objective is constructed as a robust, weighted aggregation of normalized residuals across all channels. For channel \(c\in\mathcal{S}\), the normalized residual is \(r_c(t)=[y^{\mathrm{sim}}_c(t)-y^{\mathrm{meas}}_c(t)]/s_c\), where \(s_c\) is a channel-dependent normalization factor, and separately configured for zone temperatures and return temperatures. A Huber penalty \(\rho_\delta\) is applied to reduce the influence of outliers and sporadic numerical spikes in the residuals. The total loss is then computed as
        \begin{equation}
        \mathcal{J}(\mathbf{x})=
        \frac{\sum_{c\in\mathcal{S}} w_c \sum_{t\in\mathcal{T}_c}\rho_\delta\!\big(r_c(t)\big)}
        {\sum_{c\in\mathcal{S}} w_c\,|\mathcal{T}_c|},
        \end{equation}
        where \(w_c\) is the channel weight reflecting relative importance, and \(\mathcal{T}_c\) is the set of valid time indices for channel \(c\), in which the flow-gating for return temperature channels is also included.
        
        After obtaining the total loss, the parameter update in the present paper adopts a bias-driven iterative scheme that uses the signed mean deviation of each channel. For a zone \(z\), a bias statistic is computed as
        \(b_{\mathrm{zone},z}=\mathrm{mean}(T^{\mathrm{sim}}_{\mathrm{zone},z}-T^{\mathrm{meas}}_{\mathrm{zone},z})\). For a loop \(L\), a flow-gated bias statistic is computed as
        \(b_{\mathrm{ret},L}=\mathrm{mean}(T^{\mathrm{sim}}_{\mathrm{ret},L}-T^{\mathrm{meas}}_{\mathrm{ret},L})\) evaluated only on \(\dot{m}_L\ge \dot{m}_{\min}\). These bias terms are combined into a signed direction indicator \(d_L\) that indicates whether the loop-to-zone heat exchange should be strengthened or weakened, using separate scaling for zone and return signals. The per-loop updated parameter $h_{\mathrm{int}}$ is updated in log-space with step size limiting and per-loop importance scaling by using the Python method $\mathrm{clip}$:
        \begin{equation}
        \log h_{\mathrm{int},L}^{(k+1)}=
        \log h_{\mathrm{int},L}^{(k)}+
        \mathrm{clip}\!\left(\alpha_h\,\mathrm{imp}_L\, d_L,\;-\Delta_h,\;\Delta_h\right),
        \end{equation}
        where \(\alpha_h\) is the nominal step size, \(\Delta_h\) is the maximum per-iteration log-step, and
        \(\mathrm{imp}_L\) is an error-derived importance factor that increases the update magnitude for loops with larger loss contributions. The global pipe spacing parameter $dis_{\mathrm{pip}}$ is updated synchronously based on the aggregated direction across loops:
        \begin{equation}
        \log dis_{\mathrm{pip}}^{(k+1)}=
        \log dis_{\mathrm{pip}}^{(k)}+
        \mathrm{clip}\!\left(-\alpha_d\, d_{\mathrm{g}},\;-\Delta_d,\;\Delta_d\right),
        \end{equation}
        with 
        \begin{equation}
        d_{\mathrm{g}}=\mathrm{average}(d_L;\,\mathrm{weights}=\mathrm{imp}_L),
        \end{equation}
        where the presence of the negative sign allows the pipe spacing to be reduced when the system is biased towards insufficient heating, thereby increasing effective heat transfer. To ensure physically plausible values and stable convergence, the updated parameters are constrained by predefined bounds
        \(dis_{\mathrm{pip}}\in[dis_{\min},dis_{\max}]\) and \(h_{\mathrm{int},L}\in[h_{\min},h_{\max}]\).
        \item \textbf{Damping:} A damping mechanism integrates the previous iterate with the proposed update in log-space to mitigate oscillations under strong transients:
        \begin{equation}
        \mathbf{x}_{\mathrm{use}}=(1-\lambda)\mathbf{x}^{(k)}+\lambda \mathbf{x}^{(k+1)}.
        \end{equation}
        
    \end{itemize}
    
    Finally, to improve efficiency across different channels, each channel’s loss contribution is separately computed: \(\mathrm{contrib}_c=w_c\sum_{t\in\mathcal{T}_c}\rho_\delta(r_c(t))\), and the step sizes
    \((\alpha_h,\alpha_d)\) and damping factor \(\lambda\) based on the contribution ranking are adapted to emphasize dominant error sources. This closed-loop design couples a robust objective with a physically informed update rule, allowing the calibration to progressively reduce dynamic-scale errors while preserving feasibility and numerical stability.

\section{Zoning Strategy}
\label{sec: zoning}
\begin{figure}[h]
      \centering
      \includegraphics[width=0.9\linewidth]{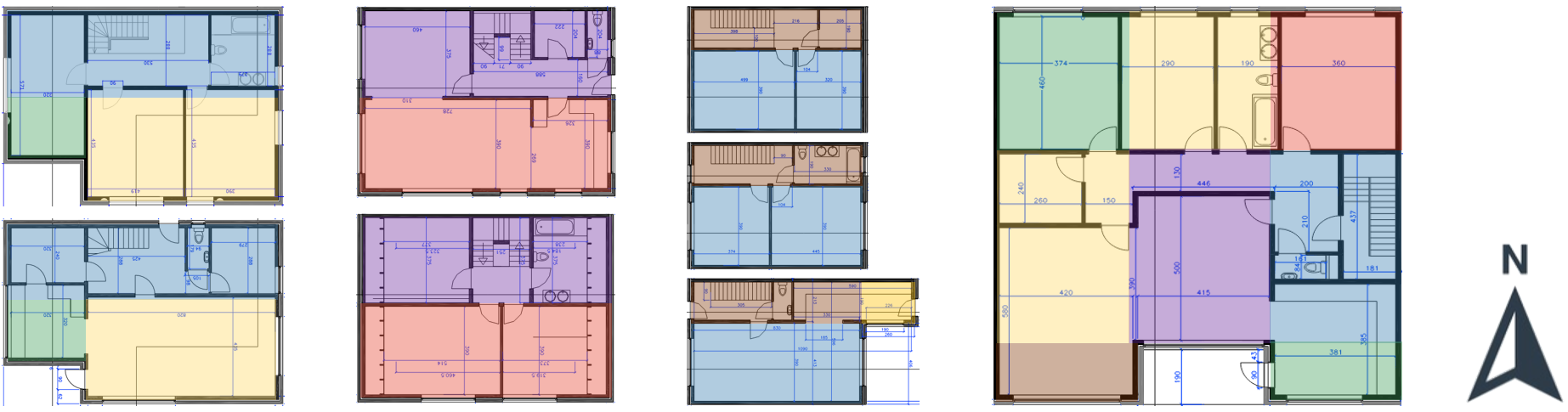}
      \caption{Zoning configurations per Annex 60 typology used in \ac{LoFi} validation (from left to right: D, S, T, A/O; color blocks denote thermal zones).}
      \label{fig: Annex zoning}  
    \end{figure}

    Zoning resolution is a key design dimension for multi-zone building energy consumption models, directly impacting physical accuracy and computational performance \citep{Pan2023}. Based on the structured zoning of building models generated using RoomFlex6D, the building is progressively subdivided into coarser and finer multi-zone configurations while maintaining consistent envelope parameters, orientation, exposure, and adjacency relationships. Figure~\ref{fig: Annex district} in Appendix~\ref{sec: appendix_annex60} shows the district studied in the Annex 60 project \citep{Wetter2017}, while Figure~\ref{fig: Annex zoning} shows architectural floor plans of five different building topologies within the district, arranged vertically from bottom to top, corresponding to the ground floor to the top floor. Each colored block represents a divided thermal zone. Further details are discussed in Section~\ref{subsec: LoFi validation}.
    
\section{Evaluation}
\label{sec: Evaluation}

    Based on the modeling tool and verification-calibration pipeline proposed in Section~\ref{sec: model construction} and~\ref{sec: V&C pipeline}, this Section first verifies the \ac{LoFi} model on the IEA EBC Annex 60 benchmark to confirm the applicability of the model generation method and zoning settings to annual and peak load indicators. Based on this, explicit hydronic loops and terminal heat exchange process are then introduced to construct a \ac{HiFi} model. Dynamic response analysis and iterative parameter calibration are carried out in combination with real-world laboratory time series measurement data, focusing on evaluating the transient consistency and improvement of the loop return water temperature and the thermal zone temperature behavior.

\subsection{Energy-Scale Validation of \ac{LoFi} Using the Annex 60 Benchmark}
\label{subsec: LoFi validation}
    The \ac{LoFi} model is first validated on the Annex 60 district benchmark using the prescribed boundary conditions and inputs, while keeping the envelope assumptions consistent across zoning configurations. For each building, (a) the annual heating demand and (b) the peak heating load are evaluated, since the former reflects energy-scale consistency, while the latter is critical for short-term and peak-oriented sector-coupled analyses. To ensure comparability with the reference study, all simulations in the present paper strictly adhere to the same boundary conditions and modeling assumptions specified in the Annex 60 final report as described in Appendix~\ref{sec: appendix_annex60}.

    Annex 60 has demonstrated that the number of thermal zones is not the primary source of differences in annual and peak heating demand across different modeling methods. Therefore, the zoning resolution of the selected zoning scheme is intentionally kept at a moderate level in Figure~\ref{fig: Annex zoning} to avoid introducing unnecessary modeling complexity. Apartments (A) and office buildings (O) are represented as single-story structures, consistent with the baseline definition. Furthermore, for buildings of the same type in different locations (e.g. Building 3 and 19 in Figure~\ref{fig: Annex district} in Appendix~\ref{sec: appendix_annex60}), the models can be fine-tuned using the orientation parameter $azi_{\mathrm{S}}$ and adjacency settings in the RoomFlex6D modeling tool to adapt to changes in energy consumption due to variations in sunlight and wind speed, ensuring that differences in simulated heating demand stem from the modeling framework, rather than differences in building definitions.

\begin{figure}[h]
      \centering
      \begin{subfigure}[t]{0.7\linewidth}
        \centering
        \includegraphics[width=\linewidth]{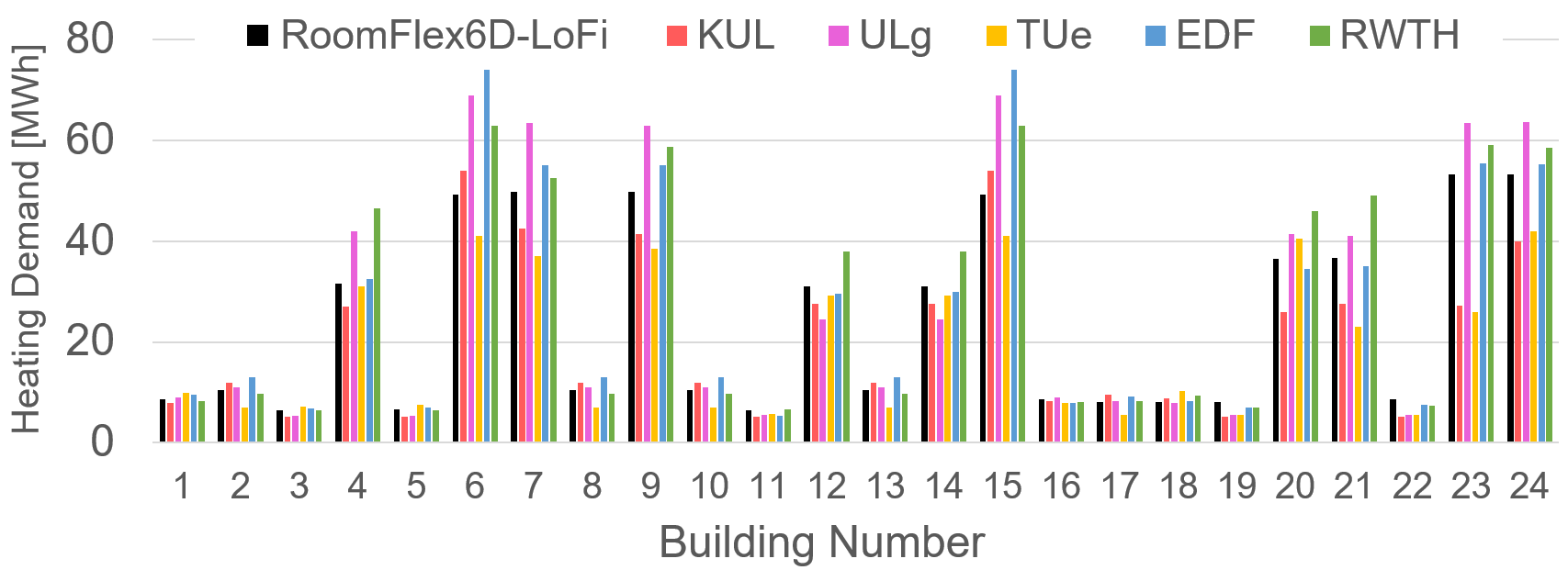}
        \caption{Comparison of \ac{LoFi} simulation results across 24 studied buildings with results from each of the 5 participants.}
        \label{fig: Annex Q_a}
      \end{subfigure}
      \begin{subfigure}[t]{0.7\linewidth}
        \centering
        \includegraphics[width=\linewidth]{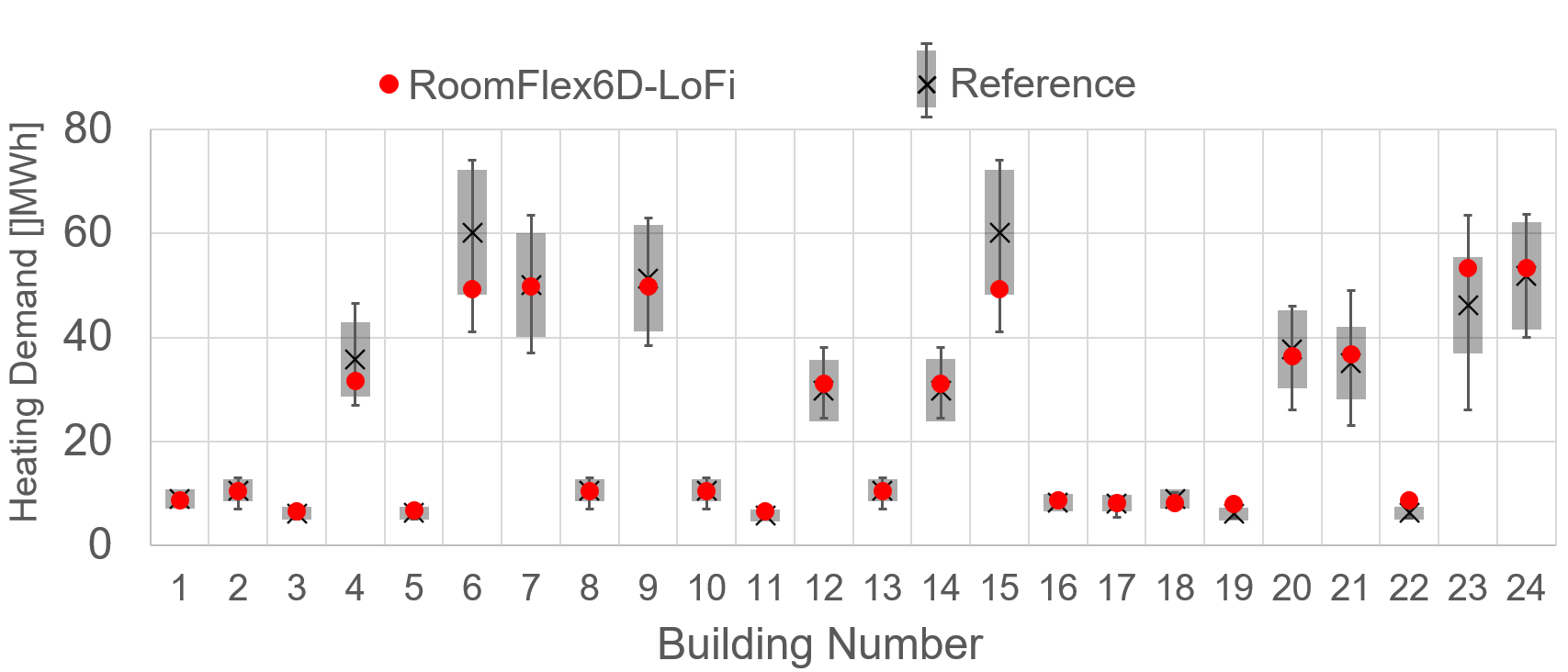}
        \caption{Comparison of \ac{LoFi} simulation results across 24 studied buildings with the average results for all participants.}
        \label{fig: Annex Q_a_ave comp}
      \end{subfigure}
      \caption{Building-level annual heating demand: \ac{LoFi} vs Annex 60 participant ranges-agreement within benchmark across 24 buildings.}
      \label{fig: Annex results Q_a}
    \end{figure}

    \textbf{Annual heating demand. }
    The energy-scale evaluation focuses on annual heating demand and peak heating load, which are the primary performance indicators defined in the Annex 60 neighborhood benchmark. Figure~\ref{fig: Annex Q_a} compares the annual heating demand of individual buildings obtained from the proposed framework with the Annex 60 reference envelope. For each building or building typology, the vertical error indicators in Figure~\ref{fig: Annex Q_a_ave comp} indicate the minimum–maximum range reported across five Annex 60 participants, the shaded bands represent errors of ±20\% based on the average of the reference values, while the red dots on the indicators represent results obtained using the presented \ac{LoFi} models. Across all buildings, the simulated annual heating demand lies within or very close to the reference error range. Differences between zoning resolutions remain limited at the annual energy scale, and no systematic overestimation or underestimation is observed for any specific building typology.

\begin{figure}[h]
      \centering
      \begin{subfigure}[t]{0.7\linewidth}
        \centering
        \includegraphics[width=\linewidth]{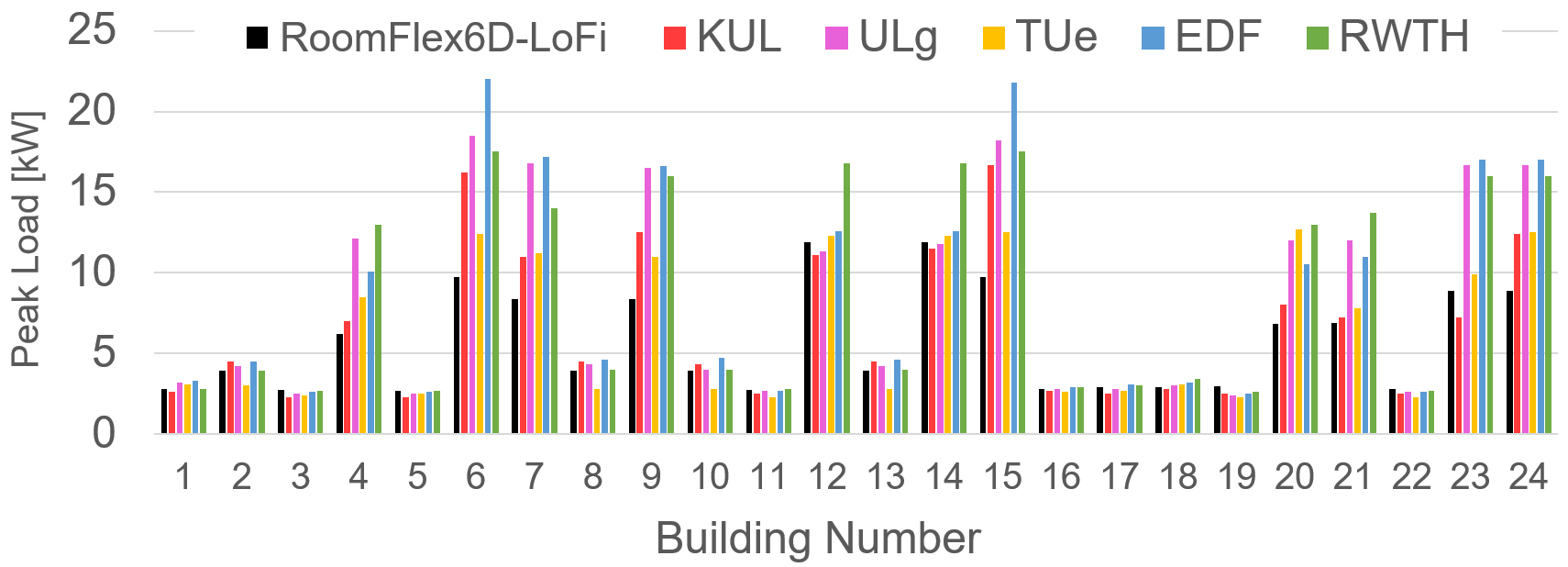}
        \caption{Comparison of \ac{LoFi} simulation results across 24 studied buildings with results from each of the 5 participants.}
        \label{fig: Annex Q_peak}
      \end{subfigure}

      \begin{subfigure}[t]{0.7\linewidth}
        \centering
        \includegraphics[width=\linewidth]{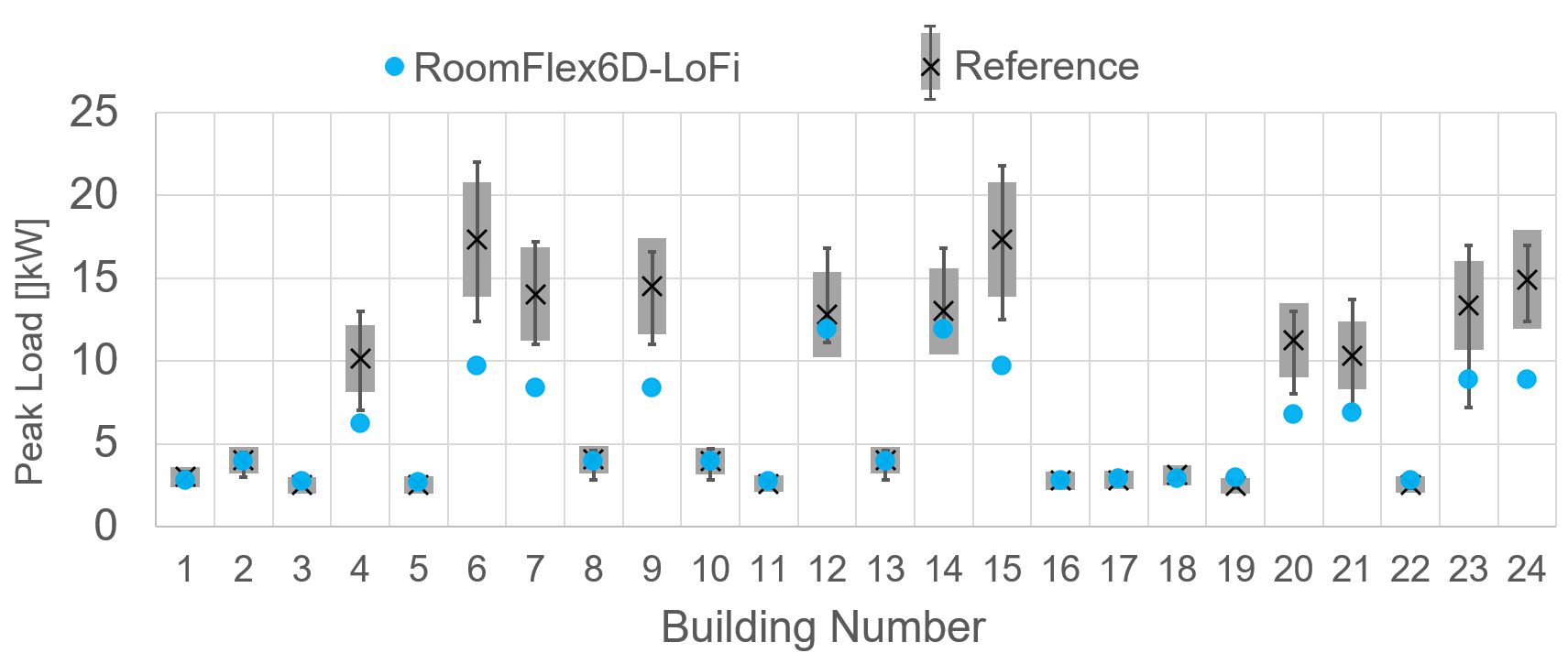}
        \caption{Comparison of \ac{LoFi} simulation results across 24 studied buildings with the average results for all participants.}
        \label{fig: Annex Q_peak_ave comp}
      \end{subfigure}
      \caption{Building-level peak heating load: \ac{LoFi} systematically underestimates high-demand peaks relative to benchmark ranges: limitation of ideal-load surrogates.}
      \label{fig: Annex results Q_peak}
    \end{figure}

    \textbf{Peak heating load. }
    While the simulation results show a high degree of agreement with the Annex 60 benchmark in terms of annual heating demand, significant deviations are observed for peak heating load, which are shown in Figure~\ref{fig: Annex results Q_peak}. In particular, for buildings with relatively high reference values, the simulated peak loads often fall below the lower limit of the Annex 60 reference range. This behavior is observed in multiple high-demand buildings, indicating a systematic underestimation of extreme peak conditions. This deviation can be explained by the inherent sensitivity of peak load indicators to short-term dynamics and modeling assumptions. In the Annex 60 benchmark, peak loads are strongly influenced by the simulation time resolution, the control strategy, and assumptions about system response under extreme outdoor conditions. In contrast, the \ac{LoFi} models used for energy-scale validation intentionally smooth fast transients and exclude explicit system-level dynamics, resulting in a more moderate peak response.

\begin{figure}[h]
      \centering
      \includegraphics[width=0.55\linewidth]{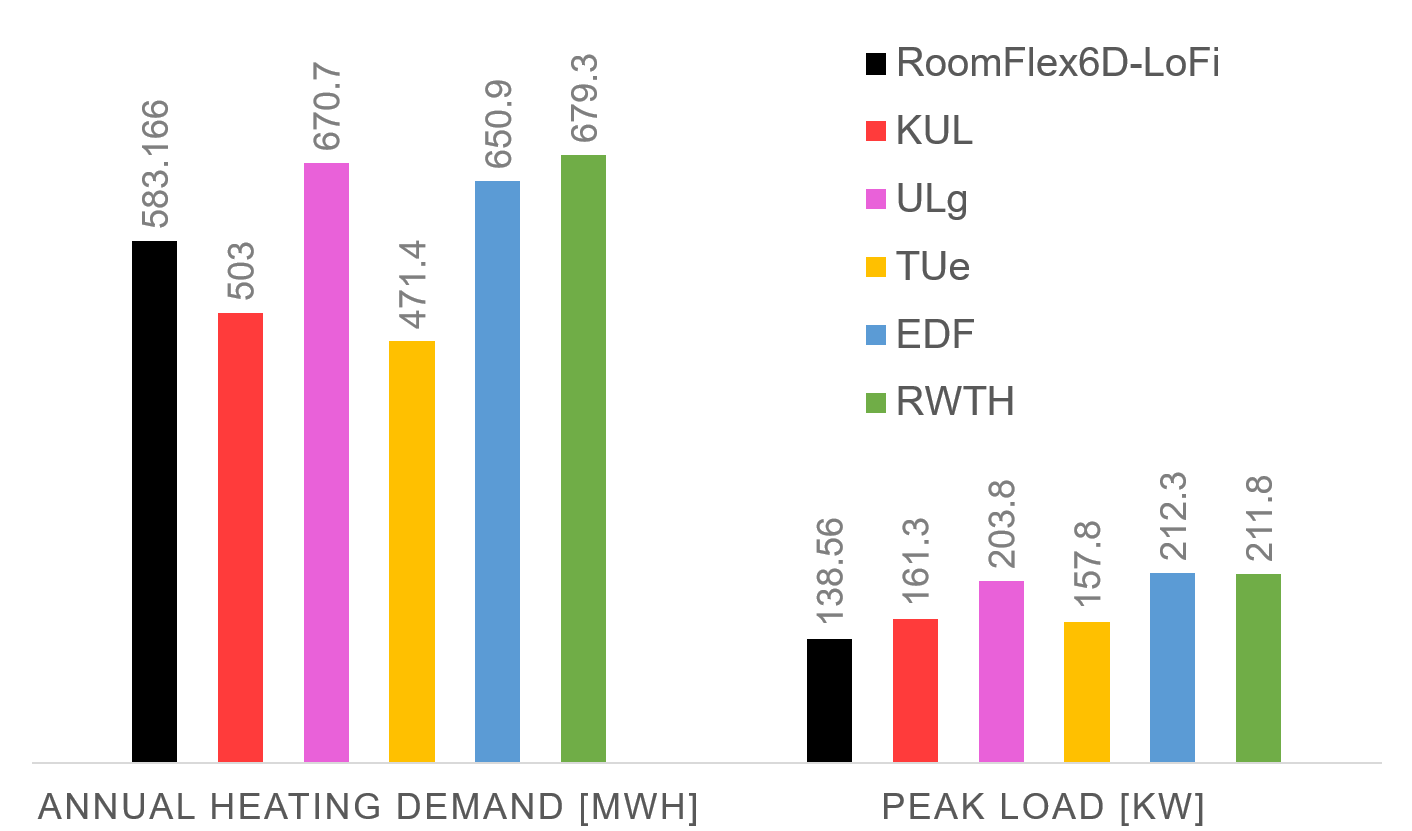}
      \caption{Neighborhood aggregation for Annex 60 benchmark: annual demand remains consistent despite peak-load underestimation at the building level.}
      \label{fig: annex overall}  
    \end{figure}

    Importantly, this underestimation does not affect the consistency of annual heating demand or the overall energy balance of the studied neighborhood, as shown in Figure~\ref{fig: annex overall}, which remains well consistent with the Annex 60 reference values. The observed differences reflect a known limitation of energy-scale surrogate models when applied to peak-oriented indices, motivating subsequent analysis using higher-fidelity models to investigate dynamic behavior and peak load characteristics in more detail.

\subsection{Dynamic-Scale Calibration of \ac{HiFi} Using Experimental Building Data}

\begin{figure}[h]
      \centering
      \includegraphics[width=0.7\linewidth]{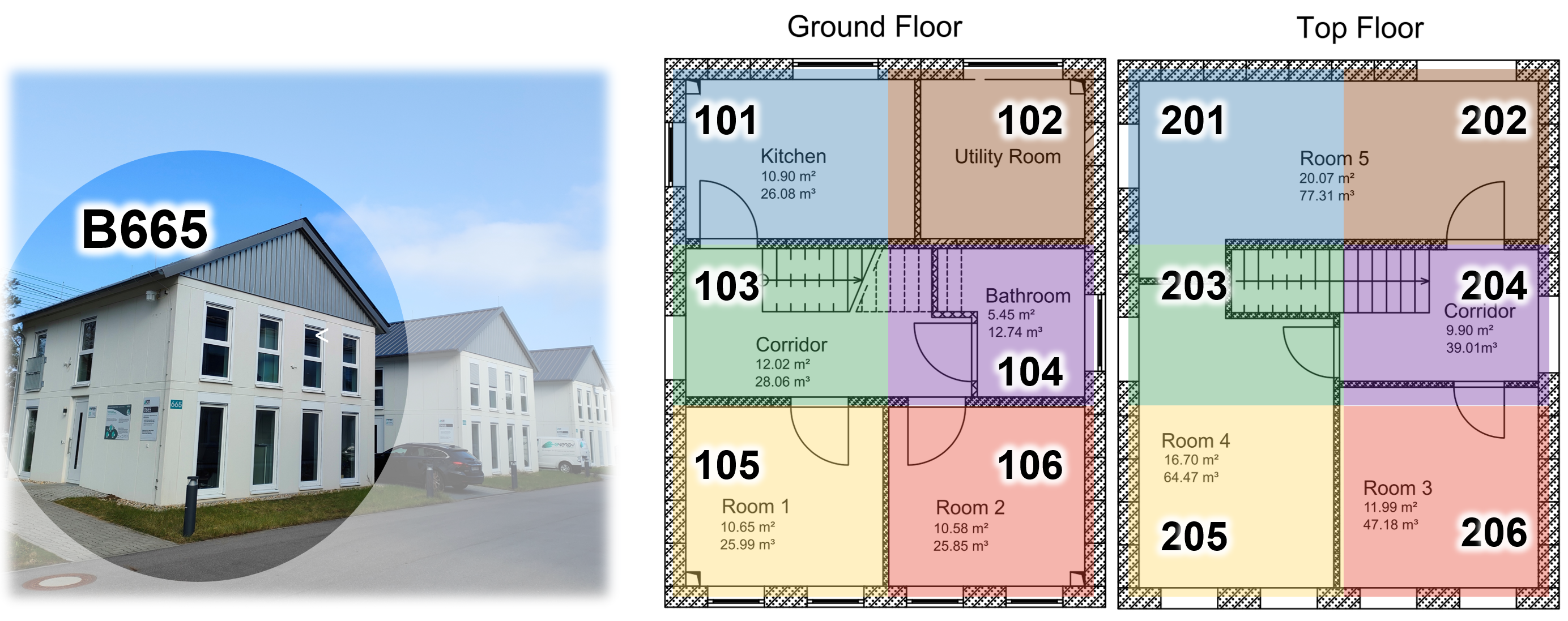}
      \caption{Living Lab \citep{Wiegel2022} B665 case: building context and 12-zone layout and IDs used for \ac{HiFi} calibration (10 UFH loops mapped to zones).}
      \label{fig: llec zoning}  
    \end{figure}

    The \ac{HiFi} case study is based on the Living Lab test building B665 \citep{Langner2025} and is implemented using the RoomFlex6D modeling tool. The building is represented using a \ac{LoFi} envelope representation, divided into a 12-zone thermal network (six zones per floor; the zoning layout and thermal zone codes are shown in Figure~\ref{fig: llec zoning}). An explicit hydronic subsystem is added, containing a total of 10 \ac{UFH} loops, that is, each thermal zone, except for zones 102 and 204, contains one \ac{UFH} loop.

    To validate and calibrate the \ac{HiFi} model, all boundary conditions, control actions, and validation objectives are taken from the Living Lab measurements over December 30, 2024 to January 3, 2025 with a 1 min sampling interval \citep{erfan2025}. The calibration focuses on two \ac{UFH}-related parameters per loop: pipe spacing distance $dis_{\mathrm{pip}}$ and the effective convective heat transfer coefficient $h_{\mathrm{int}}$ between the embedded \ac{UFH} and the zone air, as discussed in Section~\ref{sec: V&C pipeline}. In contrast, the envelope thermal parameters mainly affect the slower zone-level drifts and are difficult to identify without additional constraints. To avoid confusion and parameter non-identification, envelope parameters (e.g., ${k,d,\rho,c_p}$ from Table~\ref{tab: roomflex params}) are not tuned at this stage. Furthermore, to resolve key hydronic and control-related transients, the \ac{HiFi} model simulations employ a fixed output step size of 30 s.

\begin{figure}[h]
      \centering
      \begin{subfigure}[t]{0.33\linewidth}
        \centering
        \includegraphics[width=\linewidth]{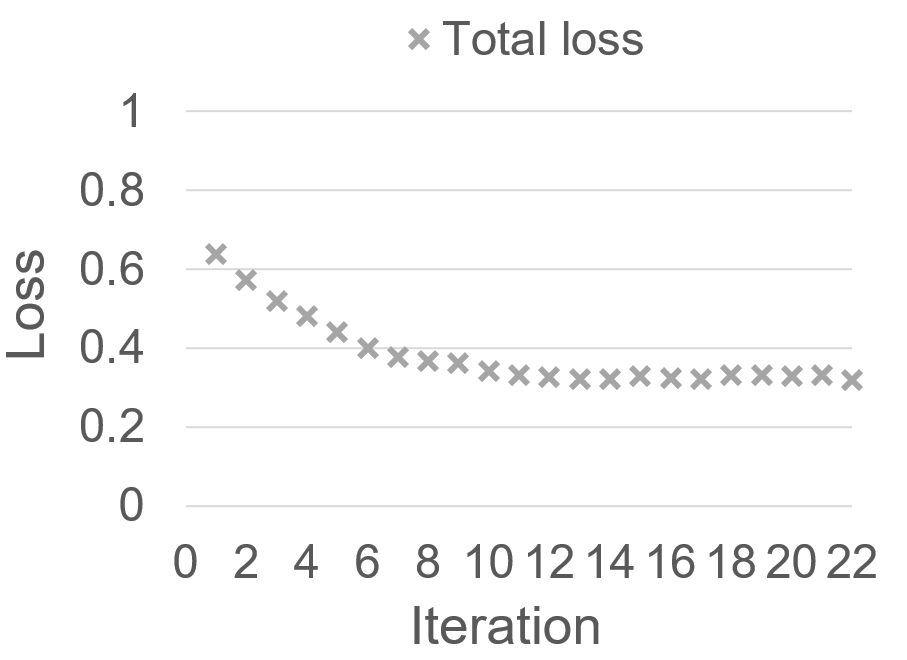}
        \caption{Total loss.}
        \label{fig: total loss}
      \end{subfigure}
      %\hfill  
      \begin{subfigure}[t]{0.42\linewidth}
        \centering
        \includegraphics[width=\linewidth]{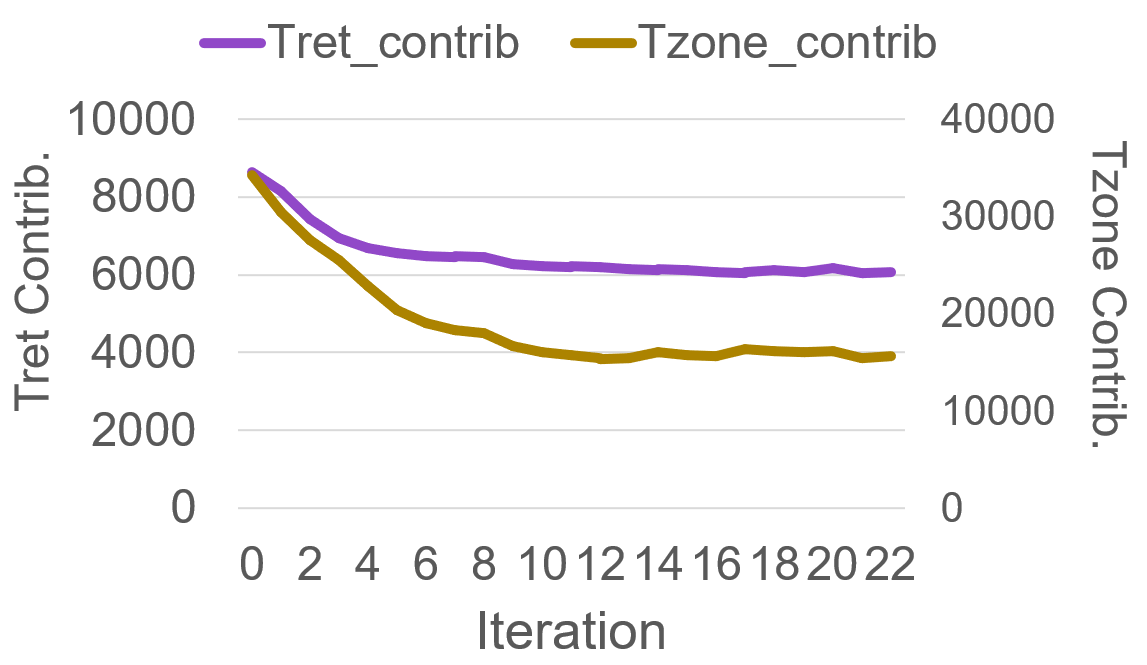}
        \caption{Loss contribution of $T_{\mathrm{ret}}$ and $T_{\mathrm{zone}}$.}
        \label{fig: loss contrib}
      \end{subfigure}
      \caption{\ac{HiFi} calibration convergence over 23 iterations: evolution of total loss and grouped contributions ($T_{\mathrm{ret}}$ gated by mass flow vs $T_{\mathrm{zone}}$). Main gains occur in the first ~8–10 iterations.}
      \label{fig: iteration loss}
    \end{figure}

    \textbf{Calibration convergence.}
    During the calibration of the \ac{HiFi} model, the authors weigh the diminishing returns against the computational cost of repeated \ac{HiFi} simulations, running 23 iterations before terminating after the target value reached significant stability: the total loss decreases from 0.64 to approximately 0.32, with the main improvement occurring in the first 8–10 iterations, followed by only minor improvements, as shown in Figure~\ref{fig: total loss}, which indicates that improvements under hydronic and thermal conditions reach a balance. It is important to note that due to the different number of signals and effective samples aggregated in the two groups of contributions (e.g., $T_{\mathrm{ret}}$ only applies when mass flow rate $\dot{m}$ > $\dot{m}_{min}$) in Figure~\ref{fig: loss contrib}, their absolute values cannot be directly compared. Nevertheless, the convergence behavior after iterations and the trend of relative loss reduction are successfully evaluated.
    
\begin{figure*}[h]
      \centering
      \begin{subfigure}[t]{0.85\linewidth}
        \centering
        \includegraphics[width=\linewidth]{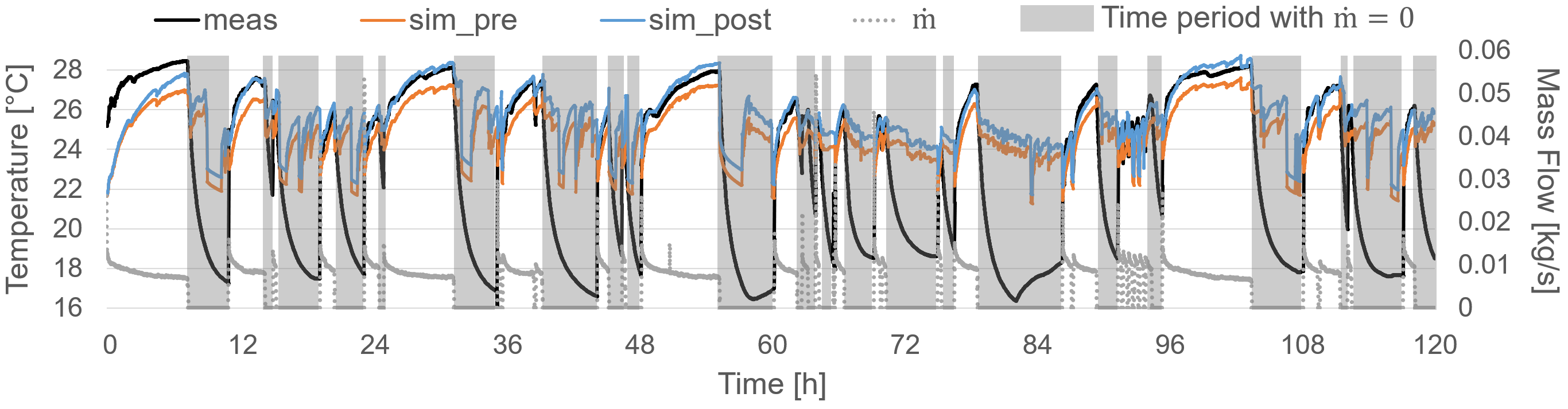}
        \caption{$T_{\mathrm{ret}}$ and $\dot{m}$ of the \ac{UFH} loop of zone 104.}
        \label{fig: Tret_104}
      \end{subfigure}
      %\hfill  
      \begin{subfigure}[t]{0.85\linewidth}
        \centering
        \includegraphics[width=\linewidth]{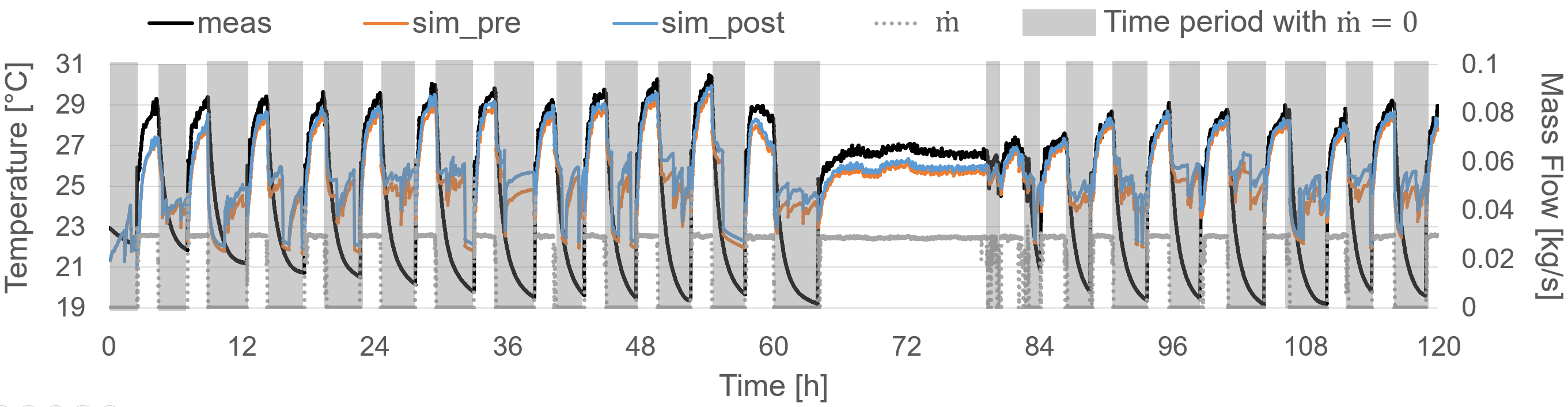}
        \caption{$T_{\mathrm{ret}}$ and $\dot{m}$ of the \ac{UFH} loop of zone 206.}
        \label{fig: Tret_206}
      \end{subfigure}
      \caption{\ac{HiFi} loop-level transient before/after calibration (zones 104 \& 206): improved ramp shape and switching response during active circulation. Gray shading indicates zero-flow intervals excluded from $T_{\mathrm{ret}}$ loss.}
      \label{fig: Tret changes}
    \end{figure*}

    \textbf{$\boldsymbol{T_{\mathrm{ret}}}$ evaluation.}
    Based on the convergence trend in Figure~\ref{fig: iteration loss}, it is evaluated whether the calibrated parameter set could improve the practically meaningful loop-level dynamics. Therefore, Figure~\ref{fig: Tret changes} presents a time-series comparison of two \ac{UFH} loops located in different zones (104 and 206) with different operating modes within the studied measurement window, where loop 104 exhibits frequent and irregular flow switching events and a significant transient ramp, while loop 206 contains regular switching events and a longer quasi-steady-state period interspersed with switching events, serving as a supplementary case for evaluating steady-state offset. In each subplot, the black curve represents the measured return water temperature $T_{\mathrm{ret}}$; the orange and blue curves represent the simulated $T_{\mathrm{ret}}$ before and after calibration, respectively. The gray-shaded area corresponds to low-flow conditions ($\dot{m}$ = 0 $kg/s$ marked by the gray dashed line on the right axis), where the return water temperature residuals are not considered in the total loss. Consistent with this logic, the largest improvements after calibration occur during periods of active circulation, where the post-calibration model better reproduces both the transient ramp shapes and the timing of return temperature responses to flow switching.

\begin{figure}[h]
      \centering
      \begin{subfigure}[t]{0.58\linewidth}
        \centering
        \includegraphics[width=\linewidth]{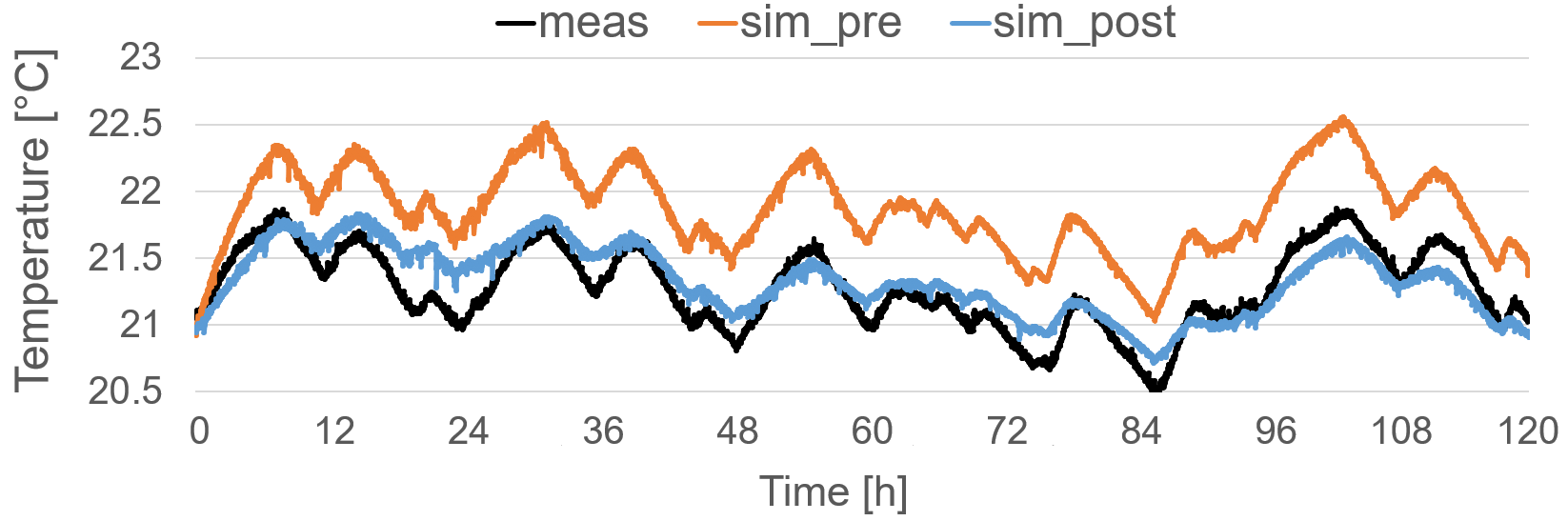}
        \caption{$T_{\mathrm{zone}}$ of zone 104.}
        \label{fig: Tzone_104}
      \end{subfigure}
      %\hfill  
      \begin{subfigure}[t]{0.58\linewidth}
        \centering
        \includegraphics[width=\linewidth]{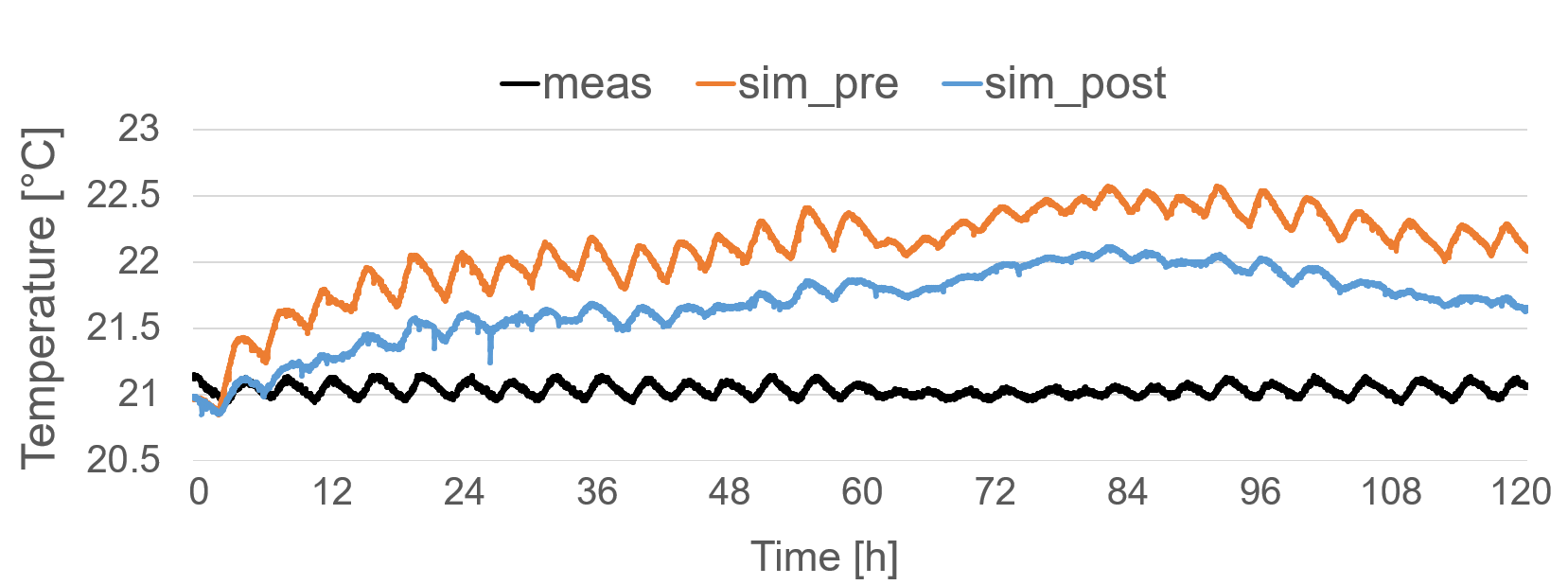}
        \caption{$T_{\mathrm{zone}}$ of zone 106.}
        \label{fig: Tzone_106}
      \end{subfigure}
      \caption{\ac{HiFi} zone-level temperatures (104 \& 106) before/after calibration: hydronic tuning improves fit without compensating errors; residual drift indicates envelope and structural uncertainty.}
      \label{fig: Tzone changes}
    \end{figure}

    \textbf{$\boldsymbol{T_{\mathrm{zone}}}$ evaluation.}
    To prevent the improved \ac{UFH} heat exchange parameters from introducing indoor temperature compensation errors, the iterative results at the zone level are evaluated in Figure~\ref{fig: Tzone changes}. To concisely and effectively examine the performance of different zones within the same calibration time window, zone 106, which shows the most significant contribution from the calibrated residual temperature, and zone 104, which operates stably and has relatively low residuals, are selected. In both zones, the calibrated trajectories are closer to the measured values, confirming that hydronic tuning does not reduce the indoor temperature fit.

    For zone 106, a residual bias of up to \SI{1.1}{\degreeCelsius} within the observation time window remains after calibration, although calibration significantly reduces the bias. The residual exhibits a slow drift across the entire time window rather than being limited to a single heating event, suggesting that the residual is primarily affected by zone-level structural uncertainties, which are weakly related to the calibrated \ac{UFH} parameters ($dis_{\mathrm{pip}}$ and $h_{\mathrm{int}}$). This may be attributed to
    \begin{itemize}
        \item \textbf{air node simplification:} the proposed model simplifies the thermal zone space into a well-mixed air node, and the sensors may measure sub-zones colder than the model's average air temperature; 
        \item \textbf{interior decorations:} actual interior modifications, such as carpets, increase the effective thermal resistance between the \ac{UFH} and the zone air, causing the model to overestimate the delivered heat;
        \item \textbf{neglected thermal mass:} unmeasured indoor thermal mass (e.g., furniture and other items) alters the effective thermal buffer and time constant, potentially resulting in an overestimation of the simulated temperature gradient; and
        \item \textbf{air infiltration:} air exchange through door and window leaks could also increase real heat losses relative to the modeled assumptions. 
    \end{itemize}
    
    Unlike the $T_{\mathrm{ret}}$-model, temperatures in thermal zones are affected by more disturbances. However, the hydronic calibration of the \ac{HiFi} model in the presented work improves loop-level dynamics and yields uncompensated improvements at the zone-level, providing guidance for future modeling improvements.

\begin{figure}[h]
      \centering
      \begin{subfigure}[t]{0.32\linewidth}
        \centering
        \includegraphics[width=\linewidth]{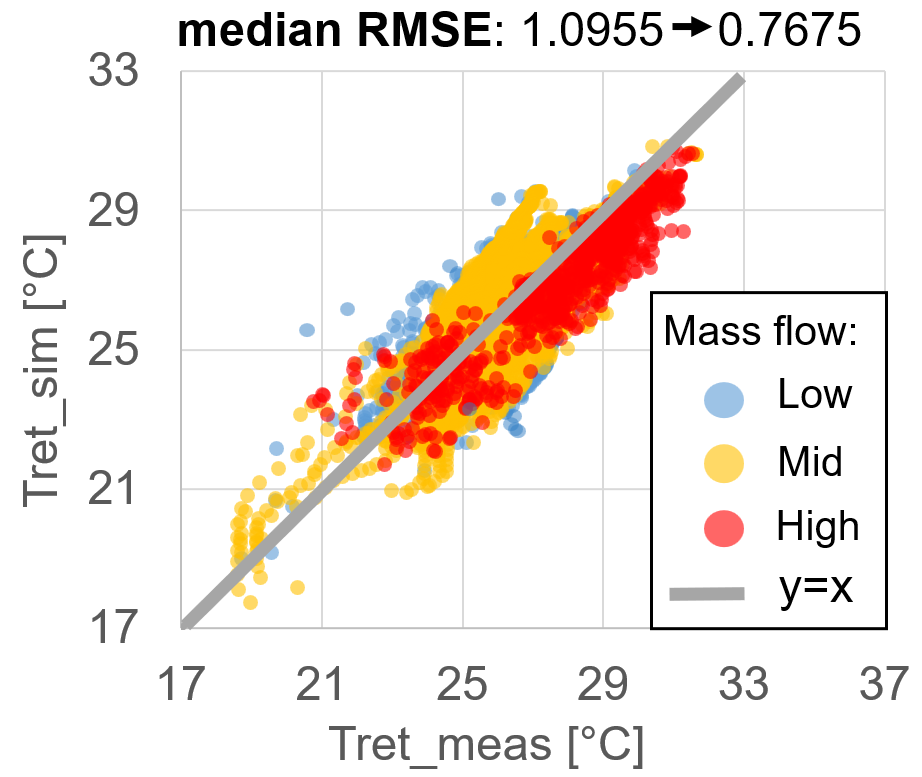}
        \caption{Scatter plot of $T_{\mathrm{ret}}$ colored by flow level.}
        \label{fig: Tret point}
      \end{subfigure}
      %\hfill  
      \begin{subfigure}[t]{0.32\linewidth}
        \centering
        \includegraphics[width=\linewidth]{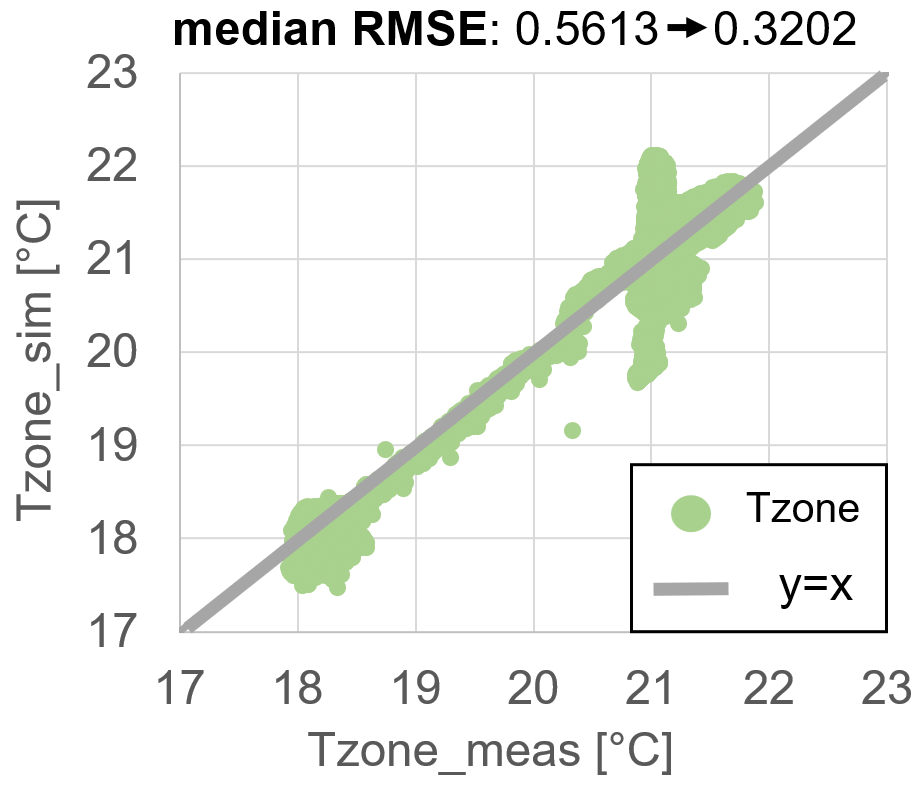}
        \caption{Scatter plot of $T_{\mathrm{zone}}$.}
        \label{fig: Tzone point}
      \end{subfigure}
      \caption{\ac{HiFi} post-calibration global consistency across all samples, identity line y=x, and median RMSE improvement.}
      \label{fig: scatter}
    \end{figure}

    \textbf{Global consistency and \ac{median RMSE}.}
    Figure~\ref{fig: scatter} supplements the convergence evidence in Figure~\ref{fig: iteration loss}, demonstrating the consistency of all hydronic loops and thermal zones on a sample-by-sample basis from a global perspective. The left subplot summarizes all aligned $T_{\mathrm{ret}}$ samples and colors them by $\dot{m}$ level, which is distinguished by \SI{0.015}{\kilo\gram\per\second} and \SI{0.03}{\kilo\gram\per\second}, thus extending the loop-level observations in Figure~\ref{fig: Tret changes} beyond the two selected loops. The scatter points are concentrated near the fitted lines (marked with grey color), while the degree of dispersion clearly depends on the operating state, that is, points at high flow rates form narrower bands, while points at low flow rates exhibit greater dispersion, which is consistent with the evaluation in Figure~\ref{fig: Tret changes}. The right subplot also summarizes all zonal temperature samples, providing a concise correspondence to the representative temperature trajectories in Figure~\ref{fig: Tzone changes}, showing that residual errors are dominated by structural uncertainties rather than insufficient convergence of the hydronic calibration. In addition to the visual consistency based on the scatter plot, Figure~\ref{fig: scatter} also reports the quantitative improvement through the \ac{median RMSE} of all signals within each group.
    
\subsection{Discussion}
\label{subsec: Discussion}
\begin{figure}[h]
      \centering
      \includegraphics[width=0.75\linewidth]{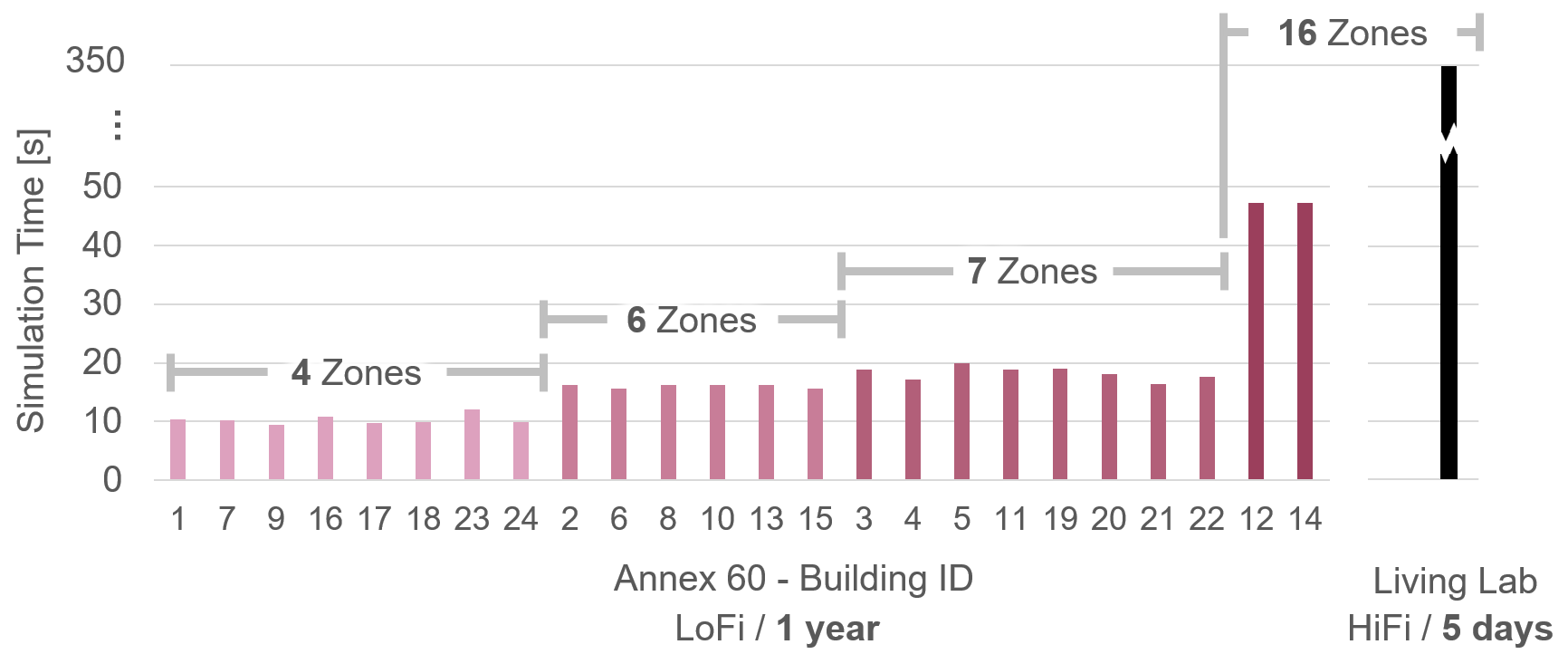}
      \caption{Runtime scaling summary: \ac{LoFi} annual Annex 60 simulations vs \ac{HiFi} 5-day Living Lab simulation: quantifying the accuracy-speed trade-off motivating fidelity choice.}
      \label{fig: sim time}  
    \end{figure}

    Increasing zonal resolution and model complexity can enhance the model's representational capabilities, but it also significantly increases simulation costs. For the \ac{LoFi} model with ideal loads, the annual simulation runtime increased from approximately \SI{10}{\second} (4-zones) to \SI{47.3}{\second} (16-zones), showing a quasi-linear increase with the number of zones (see Figure~\ref{fig: sim time} and Table~\ref{tab: t per day}). In contrast, at the same 16 zoning resolution, upgrading the \ac{LoFi} model to a high-precision \ac{HiFi} model that includes a hydronic distribution network increases the single-simulation time to approximately \SI{350}{\second} for only 5 simulated days. The main reason for this significant difference is not only the increased computational cost due to model stiffness and complexity, but also the smaller step size set in \ac{HiFi} models to achieve transient observation of dynamic behavior. Despite the large increase in cost, the \ac{HiFi} model can reproduce the dynamic changes in the loop-level return water temperature and zone-level temperature behavior under varying mass flow, which cannot be expressed by the ideal load assumption.

    \begin{table}[h]
    \centering
    \caption{Relative runtime per simulated day vs zoning and model fidelity.}
    \label{tab: t per day}
    \resizebox{0.7\columnwidth}{!}{%
    \begin{tabular}{cccccc}
    \hline
    Zoning & \ac{LoFi} (4) & \ac{LoFi} (6) & \ac{LoFi} (7) & \ac{LoFi} (16) & \ac{HiFi} (16) \\
    \hline
    Cost per simulated day & ×1.0 & ×1.6 & ×1.8 & ×4.7 & ×2555 \\
    \hline
    \end{tabular}%
    }
    \end{table}

    From an application perspective, the presented modeling framework JanusBM provides a practical guideline for modeling accuracy: 
    \begin{itemize}
        \item ideal-load \ac{LoFi} is sufficient for energy-centric studies where the decision variables depend mainly on seasonal or annual heating demand, or on average indoor temperature regulation, and where distribution constraints do not affect the results (e.g., floor-level scenario screening, studies of building renovation comparison, and long-term sector coupling planning);
        \item when temperature or power trajectories are affected by hydronic loop dynamics, hydronic \ac{HiFi} becomes necessary when required heat and deliverable heat differ due to distribution and control constraints, preferably with a zoning that maintains the loop-to-zone mapping. This includes analysis, optimization, and control design for short-term dynamics, such as peak loads and ramp rates, or studies requiring flow-related heat transfer, or loop-level performance assessments.
    \end{itemize}

\section{Conclusion and Outlook}
\label{sec: Conclusion and outlook}
    This paper presents the dual-fidelity, multi-zone white-box building modeling framework JanusBM that enables consistent analysis across annual and transient time scales by coupling a \ac{HiFi} hydronic model with a \ac{LoFi} ideal-load surrogate, both derived from the same zoning and envelope description generated by the topology-driven RoomFlex6D modeling tool. To ensure physical applicability under complementary data availability, the two-stage hybrid validation and iterative RoomFlex6D–\ac{FMU}–Python calibration pipeline using two complementary dataset is introduced. \textbf{ On the energy scale}, the \ac{LoFi} models achieve a high degree of consistency with Annex 60 in terms of annual heating demand at both building and neighborhood aggregation levels, while peak heating load is systematically underestimated in high-demand buildings. \textbf{On the dynamic scale}, the proposed iterative calibration workflow improves the transient behavior of loop-level return water temperature during active circulation periods, achieving uncompensated improvements at the zone temperature level within the same calibration window. Finally, the framework quantitatively characterizes the accuracy–speed trade-off: \ac{LoFi} simulation cost increases quasi-linearly with zoning, while introducing explicit hydronic dynamics at comparable zoning led to orders-of-magnitude higher cost, motivating a fidelity choice guided by whether required and deliverable heat are biased due to thermal distribution and control constraints. 

    Several limitations point the way for future work: First, the demonstrated peak load deviation in energy-scale \ac{LoFi} validation motivates extending the surrogate to include controlled transient capability when peak metrics are observed. Then, the current \ac{HiFi} calibration intentionally isolates hydronic impacts by tuning \ac{UFH}-related parameters while keeping building envelope parameters fixed to avoid non-identifiability, and additional constraints will be required when extending the calibration to include envelope properties. In addition, remaining structured residuals at the zone level indicate structural uncertainties beyond hydronic heat exchange, such as sensor placement, interior modifications, and air leakage, suggesting that future model extensions should account for these effects, or incorporate them with suitable grey-box representations. Finally in future work on model generation, the CityGML, IFC and gbXML source can be used as upstream providers of building geometry and attributes \citep{GMLTB2026}, which would then be compiled into the RoomFlex6D topology and parameter tables to maintain both interoperability and the efficiency required for large-scale scenario exploration.

\section*{Acknowledgments}
This work was conducted within the framework of the Helmholtz Program Energy System Design (ESD) and is partially funded under the project “Helmholtz platform for the design of robust energy systems and their supply chains” (RESUR).

\bibliographystyle{unsrtnat}
\bibliography{references}  %%% Uncomment this line and comment out the ``thebibliography'' section below to use the external .bib file (using bibtex) .





% \end{thebibliography}

\appendix

\section{RoomFlex6D \ac{GUI} Details}
\label{sec: appendix_GUI}
\begin{figure}[h]
  \centering
  \includegraphics[width=0.6\linewidth]{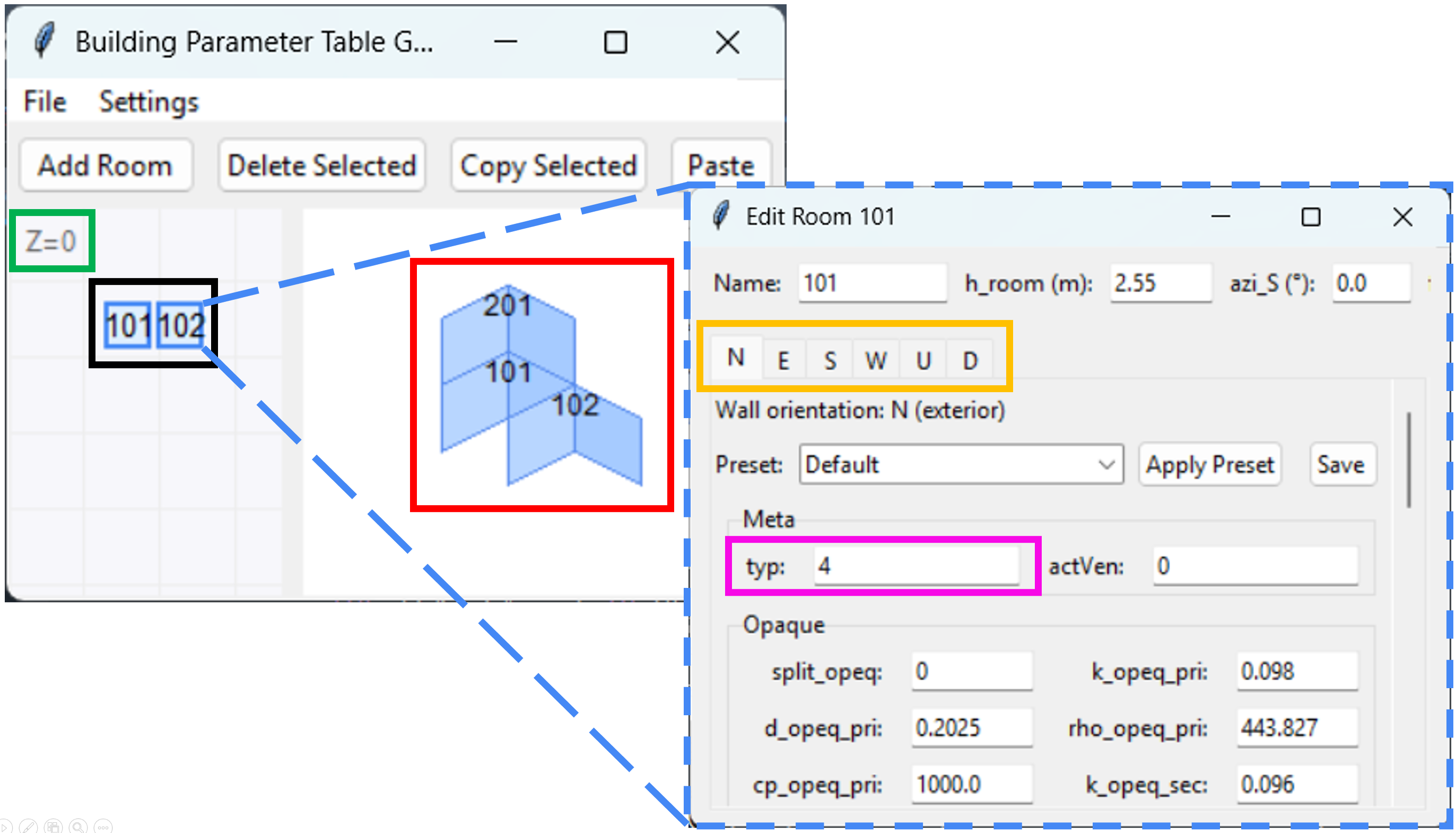}
  \caption{RoomFlex6D GUI---Further explanation for Figure~\ref{fig: GUI} with colored blocks.}
  \label{fig: GUI_total}  
\end{figure}

Figure~\ref{fig: GUI_total} illustrates the Python-based \ac{GUI} used to construct the zone topology and to edit the face-level parameter table exported by RoomFlex6D. Users need to specify adjacency relationships and boundary conditions in a structured manner before generating the Modelica model.

\subsection{Topology Configuration}
The thermal zone topology configuration page is shown on the left side of the Figure~\ref{fig: GUI_total}, where each zone is represented by an orthogonal volume (marked with a black box). The corresponding layer number of the thermal zone is displayed in the upper-left corner of the configuration page (marked with green boxes), allowing users to modify the relative placement of zones via drag-and-drop. A lightweight 3D preview area marked in red is a 3D preview, provides immediate feedback on the configured topology. 

\subsection{Zone Parameter Dialog}
By double-clicking a single thermal zone, users can configure zone-level parameters and face-level properties, as shown in the configuration window on the right side of the Figure~\ref{fig: GUI_total}. The dialog exposes the six directional faces N/E/S/W/U/D (marked with an orange box), which are the interfaces for thermal interaction with adjacent zones or the exterior.

\subsection{Face-level Boundary Specification}
The boundary conditions of each face are specified as one of five internal/external boundary types connecting to a unique adjacent area (marked with a pink box), as defined in Table~\ref{tab: adj types}, which distinguishes non-closable openings, operable windows, operable doors, opaque exterior faces, and ground-coupled exterior faces.

\subsection{Export and Model Generation}
The \ac{GUI} exports a structured face-level parameter table (shown in Table~\ref{tab: roomflex params} or Table~\ref{tab: appendix_roomflex params} in Appendix~\ref{sec: appendix_Parameter table}) that is used for the subsequent automatic Modelica model generation (Figure~\ref{fig: example mo building} for an Example). Therefore, updating the topology or parameters only requires regenerating the table and re-instantiating the Modelica model, supporting fast and repeatable zoning and structural modifications.

\section{\ac{HiFi} Hydronic Subsystem Specification and Configuration in RoomFlex6D}
\label{sec: appendix_HiFi}
RoomFlex6D allows users to configure a building hydronic heating subsystem and couple it to the generated multi-zone envelope structure via the structured parameter table in Table~\ref{tab: roomflex params} or Table~\ref{tab: appendix_roomflex params} in Appendix~\ref{sec: appendix_Parameter table}. The \ac{HiFi} model addresses transient and control-relevant behaviors by explicitly representing both the heater-side thermodynamics and the network-side hydronic dynamics.

\subsection{Heater Models}
Radiator and \ac{UFH} systems are supported as typical configurations in the proposed framework JanusBM. In both cases, heaters are considered as dynamic components with finite heat capacity. The heat exchange of a building is modeled through convection and radiation to the zone air, or through embedded constructions in the case of \ac{UFH}, which enables delayed heat release effects caused by construction thermal inertia.

\subsection{Hydronic Network Components}
The network comprises supply and return pipes, hydraulic flow resistances, circulation pumps, and control elements including control valves, bypass, and mixing valves. Mass flow rate and pressure drop are computed based on conservation laws by the Modelica algorithm. Meanwhile, the state variables include the supply water temperature and the opening state of control valves, which can be driven by zone-temperature feedback within a closed-loop control, or prescribed as external inputs to represent scenario- or controller-dependent operation.

\subsection{Configuration Parameters by the GUI}
The \ac{GUI} supports configuring (a) heater type (radiator vs \ac{UFH}), (b) the loop structure and the sequence through which hot water flows across heaters, and (c) \ac{UFH}-specific parameters such as embedding depth and other settings that influence heating dynamics. These settings are exported in the hydronic-related columns of Table~\ref{tab: roomflex params} or Table~\ref{tab: appendix_roomflex params} in Appendix~\ref{sec: appendix_Parameter table}, including heater type, loop identifier, flow order, and control mapping, providing a reproducible interface between topology definition and Modelica model generation.

\section{Background Annex 60 Benchmark}
\label{sec: appendix_annex60}
\begin{figure}[h]
    \centering
    \includegraphics[width=0.5\textwidth]{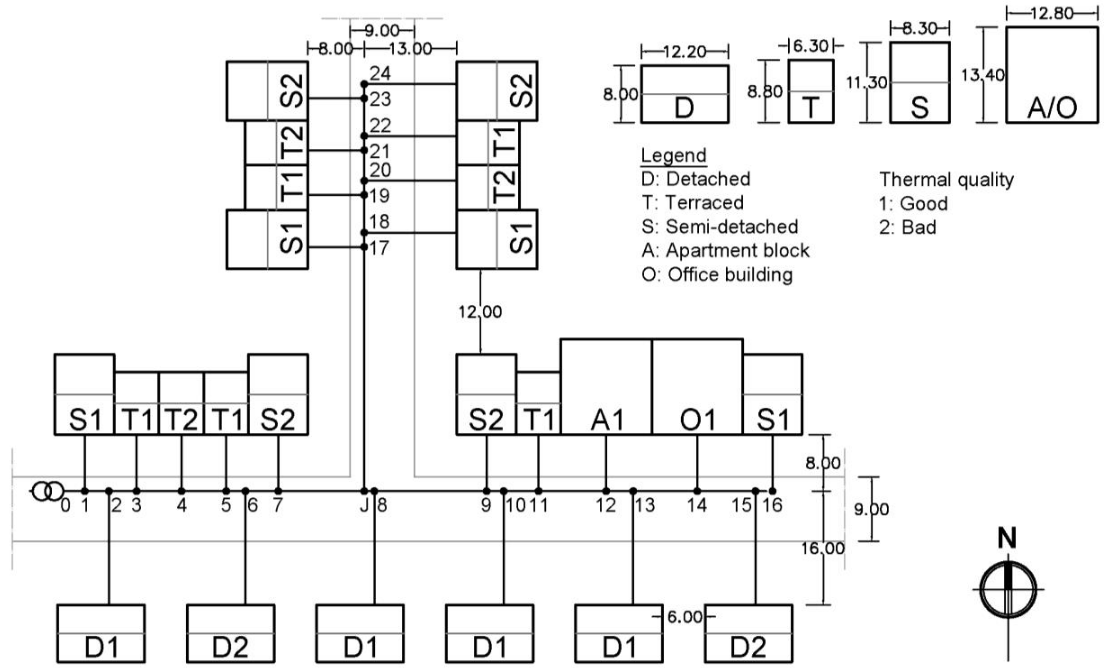}
    \caption{Annex 60 district layout used for \ac{LoFi} validation (24 buildings, five typologies).}
    \label{fig: Annex district}
\end{figure}

The studied Annex 60 neighborhood (shown in Figure~\ref{fig: Annex district}) consists of 24 buildings with five building types: detached houses (D), semi-detached buildings (S), terraced buildings (T), apartment (A), and office buildings (O), where two different thermal insulation stages of building envelopes were represented by number 1 (better) and 2 (worse). Five modeling teams (TUe, EDF, KUL, ULg and RWTH Aachen) in Annex 60 benchmark were required to use the Modelica modeling language and adhere the same boundary conditions for building simulation, including:
\begin{itemize}
    \item a reference climate in Uccle, Belgium,
    \item a constant heating setpoint of 21°C,
    \item zero internal heat gain,
    \item no shading, and
    \item a baseline-defined heating season (from October 10th to May 19th of the following year).
\end{itemize}

Nevertheless, these teams had the flexibility to choose their own modeling methods regarding model complexity and libraries. Zoning strategies of these participants ranged from single-zone representations to multi-zone configurations with one or more zones per floor, depending on building type and facade orientation. Some participants explicitly simulated heat transfer between adjacent buildings, while others neglected inter-building thermal coupling. This diversity in modeling choices resulted in significant differences in the predicted annual and peak heating demands, as described in the final report of Annex 60. 

Despite these unified boundary conditions, Annex 60 was reported as participant result ranges rather than a unique reference value. Common sources of variation include (but are not limited to):
\begin{itemize}
    \item \textbf{inter-building coupling assumptions:} whether and how thermal interactions with adjacent buildings or shared boundaries are represented;
    \item \textbf{envelope and boundary idealizations:} e.g., discretization of envelope elements, treatment of thermal bridges, and the handling of ground-coupled surfaces and adjacent areas; and
    \item \textbf{numerical settings and solver-related effects:} the settings of time step and convergence tolerances can influence peak metrics more strongly than annual energy metrics.
\end{itemize}

\section{Parameter Table Structured by RoomFlex6D Modeling Tool}
\label{sec: appendix_Parameter table}
\begin{center}
\rotatebox{90}{%
    \begin{minipage}{0.83\textheight}
        \centering
        \captionof{table}{Full version of face-level parameter-table schema generated by RoomFlex6D for model generation.}
        \label{tab: appendix_roomflex params}
        \renewcommand{\arraystretch}{1.1}
        \begin{tabularx}{0.83\textheight}{l c X c c}
            \hline
            \textbf{Parameter} 
            & \textbf{Symbol} 
            & \textbf{Description} 
            & \textbf{Type} 
            & \textbf{Example} \\
            \hline
            Primary zone ID      
            & $z_{\mathrm{pri}}$ 
            & Index of the current thermal zone 
            & Integer 
            & 101 \\
            
            Face orientation     
            & $ori$              
            & Face direction (1--6 corresponding to N/E/S/W/U/D) 
            & Integer 
            & 2 \\
            
            Adjacent zone ID     
            & $z_{\mathrm{adj}}$ 
            & Identifier of the neighboring zone (0 denotes an external boundary) 
            & Integer 
            & 102 \\
            
            Boundary type        
            & $typ$              
            & (See Table~\ref{tab: adj types})
            & Integer 
            & 3 \\
            
            Ventilation flag     
            & $actVen$           
            & Indicates whether active ventilation is applied on this face 
            & Integer 
            & 0 \\
            
            Split flag   
            & $split$     
            & Indicates whether \ac{UFH} is NOT installed
            & Integer 
            & 1 \\
            
            Thermal parameter  
            & ${k,d,\rho,c_p,A}_{\mathrm{opeq,fra,gla,open,RS}}$ 
            & Thermodynamic parameters and area of walls, door and window frames, glass, openings, and roller shutters. 
            & Real 
            & 0.70 \\
        
            Hydronic  
            & $heat_{\mathrm{type,loop,order,ctrl}}$ 
            & Types, numbering, sequence, and control schemes of heating loops. 
            & Integer 
            & 1 \\
        
            Building orientation
            & $azi_{\mathrm{S}}$ 
            & The angle in degrees between the user-defined south-facing surface and true south, used to calculate the impact of solar radiation and wind speed.
            & Real 
            & 2.3 \\
        
            \hline
        \end{tabularx}
    \end{minipage}%
  }
\end{center}

\end{document}